\RequirePackage{fix-cm}
\RequirePackage{fix-cm}
\documentclass[superscriptaddress,amssymb,amsmath,nobibnotes,showkeys,showpacs,nofootinbib,smallextended]{svjour3}
\smartqed  
\RequirePackage{graphicx}
\usepackage{graphicx}
\usepackage{placeins}
\usepackage{float}

\RequirePackage{latexsym}
\RequirePackage[numbers,sort&compress]{natbib}
\RequirePackage[colorlinks,citecolor=blue,urlcolor=blue,linkcolor=blue]{hyperref}
\journalname{Eur. Phys. J. C}
\usepackage{graphicx}
\usepackage{dcolumn}
\usepackage{bm}
\usepackage[
scale=0.7, marginratio={1:1, 2:3}, ignoreall,
text={7in,10in},centering,
margin=1.5in,
total={6.5in,8.75in}, top=1.2in, left=0.9in, includefoot,
height=10in,a5paper,hmargin={3cm,0.8in},
]{geometry}
\usepackage{widetext}
\RequirePackage{graphicx}
\RequirePackage{mathptmx}      
\usepackage{adjustbox}
\RequirePackage{latexsym}
\RequirePackage[colorlinks,citecolor=blue,urlcolor=blue,linkcolor=blue]{hyperref}
\usepackage[english]{babel}
\usepackage{amssymb,amsmath,amsfonts,graphicx, graphics, setspace}
\usepackage{bigints}
\usepackage{anysize}
\usepackage{float}
\usepackage{fancyhdr}
\usepackage{subfigure}
\usepackage{pdflscape}
\usepackage{epsfig}
\usepackage{epsf}
\usepackage{epstopdf}
\usepackage{hyperref} 
\usepackage{amsmath}
\usepackage{amsfonts}
\usepackage{amssymb}
\usepackage{subfigure}
\usepackage{epstopdf}
\usepackage{epsf}
\usepackage{url} 
\usepackage{pdflscape} 
\usepackage{amsmath}
\usepackage{amssymb} 
\usepackage{epsfig} 
\usepackage{amsfonts}
\usepackage{graphicx}
\usepackage{times,fancyhdr}
\usepackage{enumitem}
\setlist[enumerate,2]{label=\roman*)}

\def\case#1/#2{\textstyle\frac{#1}{#2}}

\newcommand{\be}{\begin{equation}}
\newcommand{\ee}{\end{equation}}
\newcommand{\ben}{\begin{eqnarray}}
\newcommand{\een}{\end{eqnarray}}

\usepackage{amsmath}
\usepackage{amsfonts}
\usepackage{amssymb}
\usepackage{widetext}
\usepackage{epstopdf}
\usepackage{epsf}

\begin{document}

\title{Inflation Driven by Non-Linear Electrodynamics}

\author{H. B. Benaoum \and Genly Leon \and A. {\"O}vg{\"u}n   \and H. Quevedo}
\institute{H. B. Benaoum \at  Department of Applied Physics and Astronomy, University of Sharjah, United Arab Emirates \email{hbenaoum@sharjah.ac.ae} \and Genly Leon \at Departamento  de  Matem\'aticas,  Universidad  Cat\'olica  del  Norte, Avda. Angamos  0610,  Casilla  1280  Antofagasta,  Chile  \at  Institute of Systems Science, Durban University of Technology, PO Box 1334, Durban 4000, South Africa \email{genly.leon@ucn.cl} \and 
 A. {\"O}vg{\"u}n \at Physics Department, Eastern Mediterranean University, Famagusta, 99628 North Cyprus via Mersin 10, Turkey
\email{ali.ovgun@emu.edu.tr}
\url{https://aovgun.weebly.com/} 
\and  H. Quevedo  \at Instituto de Ciencias Nucleares, Universidad Nacional Aut\'onoma de M\'exico,
 AP 70543, Ciudad de M\'exico 04510, Mexico  \at  Dipartimento di Fisica and ICRANet, Universit\`a di Roma ``La Sapienza",  I-00185 Roma, Italy\email{quevedo@nucleares.unam.mx}
}

\date{\today}

\maketitle
\begin{abstract}{We investigate the inflation driven by a nonlinear electromagnetic field based on an NLED lagrangian density ${\cal L}_{\text{nled}} = - {F} f \left( {F} \right)$, where $f \left( {F}\right)$ is a general function depending on ${F}$. We first formulate an $f$-NLED cosmological model with a more general function $f \left( {F}\right)$ and show that all NLED models can be expressed in this framework; then, we investigate in detail two interesting examples of the function $f \left( {F}\right)$. We present our phenomenological model based on a new Lagrangian for NLED. Solutions to the field equations with the physical properties of the cosmological parameters are obtained. We show that the early Universe had no Big-Bang singularity, which accelerated in the past. We also investigate the qualitative implications of NLED by studying the inflationary parameters, like the slow-roll parameters, spectral index $n_s$, and tensor-to-scalar ratio $r$, and compare our results with observational data. Detailed phase-space analysis of our NLED cosmological model is performed with and without matter source. As a first approach, we consider the motion of a particle of unit mass in an effective potential. Our systems correspond to fast-slow systems for physical values of the electromagnetic field and the energy densities at the end of inflation. We analyze a complementary system using Hubble-normalized variables to investigate the cosmological evolution before the matter-dominated Universe. }
\end{abstract}



\section{Introduction} 
The inflationary paradigm \cite{starobinsky1980new,guth1981,linde1982,albrecht1982,abbott984,lucchin1985} of the early Universe has become a crucial part of the standard cosmological model since it has received tremendous support from the latest observational data \cite{ade2015, Planck2018,akrami2018,ade2015b,akrami2020}. 
According to this scenario, the early Universe underwent an accelerated expansion that could solve several puzzles of the hot Big Bang cosmology, such as the flatness problem and the horizon problem, and provide a mechanism to generate primordial cosmological perturbations \cite{Mukhanov:1981xt}. The most straightforward approach to describe the inflationary era is to use a canonical scalar field with self-interacting potential. A variety of inflationary models have been proposed, such as non-minimal Higgs inflation \cite{shaposhnikov2008}, Starobinsky inflation \cite{starobinsky1980new} and {\ colour {red} others (see \cite{martin2014encyclopaedia})}.
Despite the impressive success of inflation, the standard cosmological model has a cosmological singularity at a finite time in the past where the curvature and energy density are not finite \cite{caroll2001, hawking1966}. Some proposals of cosmological models are free of any singularities based on various distinct mechanisms. For an incomplete list of non-singular cosmological models, see \citep{muk, novello,saless1990,brandenberger1998,cai2012,trodden1993}.

Another approach developed by Born and Infeld (BI) \cite{BIa, BIb, BIc} as a way to cure the divergences of self-energy of charged particles is to replace the original Maxwell Lagrangian with a nonlinear electrodynamics (NLED) Lagrangian. Similarly, Plebanski studied different models of NLED Lagrangians and proved that the BI model satisfies physically acceptable requirements \cite{Gutierrez:1981ed}. There are various applications of NLED in the literature, including cosmology and astrophysics \cite{MosqueraCuesta:2009tf, Corda:2010ni, MosqueraCuesta:2017iln, Baccigalupi:2002bh, MosqueraCuesta:2003dh, MosqueraCuesta:2004wh, Mbelek:2006mp, Mbelek:2007um}, high power laser technology, plasma physics, nonlinear optics \cite{Lundin:2006xn, Marklund:2006my, Delphenich:2006ec, Lundstrom:2005za} and the field nonlinear exponential growth due to chiral plasma instability \cite{Akamatsu:2013pjd}. In this framework, the standard cosmological model based on Friedmann-Lema{\^i}tre-Robertson-Walker (FLRW) geometry with the nonlinear electromagnetic field as its source leads to a cosmological model without primordial singularity. Other interesting NLED models have been introduced in the literature \cite{kruglov1,kruglov2,kruglov3,ao,sadia,magneticU1,sound,novello1,novello2,vol1,novello3,novello4,novello5,w,w1,Campanelli:2007cg,DeLorenci,kruglov4,novello6,Aiello:2008,10,Montiel:2014dia,Ovgun:2017iwg,GarciaSalcedo:2002jm,kruglo5,kruglov6,hbenaoum2021}. 

In \cite{hbenaoum2021}, Benaoum and {\"O}vg{\"u}n have proposed a phenomenologically viable cosmological model based on NLED that could address some open cosmological problems such as the absence of primordial singularity, an early acceleration of the Universe, and the generation of matter-antimatter. One of the exciting features of nonlinearity is the removal of the initial singularity. We have assumed that a stochastic magnetic field background fills the Universe. 
Magnetic fields are believed to have played a crucial role in the evolution of the Universe, and it is not surprising that our Universe is teeming with magnetic fields. Magnetic fields are everywhere in our Universe \cite{neronov2010,taylor2011}. 
Constraints on the magnetic fields depend on the generation mechanism of the primordial magnetic field \cite{paoletti2022constraints,paoletti2019constraints}. 
In particular, a lower bound on the strength of the magnetic field of the order of $B \geq 3 \times 10^{-16}~G$ has been obtained for
intergalactic magnetic fields \cite{ade2016a} whereas the Planck satellite in 2015 gives an upper limit to be of the order of $B < 10^{-9}~G$ \cite{anwar2015}.
However, very little is known about the existence and origin of magnetic fields in the early Universe \cite{grasso2001,subramanian2016, MosqueraCuesta:2011tz}. Finding primordial magnetic fields would transform our understanding of how our Universe evolved. 

Our main aim is to study a new generalized case of NLED Lagrangian density which can be important in the very early Universe, leading to the avoidance of the singularity. To do so, in the present work, we investigate the inflation driven by a nonlinear electromagnetic field based on an NLED lagrangian density ${\cal L}_{\text{nled}} = - {F} f \left( {F} \right)$, where $f \left( {F}\right)$ is a general function of ${F}$. The nonlinearity is encoded in the function $f \left( {F}\right)$. We first formulate an $f$-NLED cosmological model with a more general function $f \left( {F}\right)$ and show that all NLED models can be expressed in this framework; then, we investigate in detail two interesting examples of the function $f \left( {F}\right)$. The outline of the paper is as follows. In section \ref{sec:nled}, we present our phenomenological model based on a new Lagrangian for NLED. Solutions to the field equations with the physical properties of the cosmological parameters are obtained. Here, we show that the early Universe had no Big-Bang singularity and tended to accelerate in the past. We also investigate the qualitative implications of NLED by studying the inflationary parameters, like the slow-roll parameters, spectral index $n_s$, and tensor-to-scalar ratio $r$, and compare our results with observational data. Detailed phase-space analysis of our NLED cosmological model is performed in 
section \ref{sec:phase} with and without matter source. Finally, we devote section \ref{sec:con} to our conclusions.

\section{General Relativity Coupled to Nonlinear Electrodynamics} 
\label{sec:nled}
The action of Einstein's gravity coupled with NLED is given as follows: 
\begin{eqnarray}
S & = & \int d^4 x \sqrt{-g} \left( \frac{1}{2} R + {\cal L}_{\text{nled}} \right), 
\label{eq1}
\end{eqnarray}
where $R$ is the Ricci scalar, ${\cal L}_{\text{nled}}$ is the NLED Lagrangian, and we use geometrized units, where $8\pi G=1, c=1$.

In general, the NLED Lagrangian can be expressed as a function of ${F} = \frac{1}{4} F_{\mu \nu} F^{\mu \nu}$ and ${\cal G} = \frac{1}{4} F_{\mu \nu} \widetilde{F}^{\mu \nu}$, where $F_{\mu \nu}$ is the field strength tensor and $\widetilde{F}_{\mu \nu}$ is dual. 
Since the classical Maxwell theory is valid in the low-energy/weak-coupling limit, the NLED Lagrangian reduces to the Maxwell one, i.e. ${\cal L} = - {F}$ in the corresponding limit. 
Here, we restrict ourselves to the case of an NLED Lagrangian depending on the electromagnetic field strength 
invariant ${F}$ where the classical Maxwell's Lagrangian density is replaced by
\begin{eqnarray}
{\cal L}_{\text{nled}} & = & - {F} f \left( {F} \right),
\label{eq2}
\end{eqnarray}
where $f \equiv f \left( {F}\right)$ is a general function depending on ${F}$.  

Variation of the action for the metric and the NLED fields leads to  the following field equations,  
\begin{equation}
R_{\mu \nu }-\frac{1}{2}g_{\mu \nu }R=  T_{\mu \nu },  \label{eq3}
\end{equation}
and $\ $%
\begin{equation}
\partial _{\mu }\left( \sqrt{-g}\frac{\partial \mathcal{L}_{\text{nled}}}{\partial F }F^{\mu \nu }\right) =0.
\label{eq4}
\end{equation}
where $T_{\mu \nu}$ is the energy-momentum tensor of the NLED fields,
\begin{eqnarray}
T^{\mu \nu} = H^{\mu \lambda} F^\nu_{\ \lambda} - g^{\mu \nu} {\cal L}_{\text{nled}},~~~~
H^{\mu \lambda} =  \frac{\partial {\cal L}_{\text{nled}}}{\partial F_{\mu \lambda}} = \frac{\partial {\cal L}_{\text{nled}}}{\partial {F}} F^{\mu \lambda}.
\label{eq5}
\end{eqnarray}

From the above NLED Lagrangian density, the energy-momentum tensor can be written as: 
\begin{eqnarray}
T^{\mu \nu} & = & - \left( f + {F} f_{{F}} \right)  F^{\mu \lambda} F^{\nu}_{\ \lambda} + g^{\mu \nu} {F} f,
\label{eq6}
\end{eqnarray}
where $f_{{F}}=\frac{d f}{{d F}}$. The energy density $\rho$ and pressure $p$ can be obtained as follows:
\begin{eqnarray}
\rho_{\text{nled}} & = & {F} f - E^2 \left( f + {F} f_{{F}} \right), \nonumber \\
p_{\text{nled}} & = & - {F} f + \frac{2 B^2 - E^2}{3} \left(f + {F} f_{{F}}  \right).
\label{eq7}
\end{eqnarray}
Assuming that the stochastic magnetic fields are the cosmic background with a wavelength smaller than the curvature, we can use the averaging of EM fields which are sources in GR, to obtain an FLRW isotropic spacetime \cite{tolman}. The averaged EM fields are as follows: 
\begin{equation}
\langle \mathbf{E}\rangle =\langle \mathbf{B}\rangle =0,\text{ }\langle E_{i}B_{j}\rangle =0, 
\label{eq8}
\end{equation}
\begin{equation*}
\langle E_{i}E_{j}\rangle =\frac{1}{3}E^{2}g_{ij},\text{ }\langle B_{i}B_{j}\rangle =\frac{1}{3} B^{2} g_{ij}.
\end{equation*}
where the averaging brackets $\langle $ $\rangle $ is used for simplicity.  

In what follows, we consider the case where the electric field vanishes, i.e. $E^2 = 0$,  
and a non-zero averaged magnetic field leads to a magnetic Universe. Such a purely magnetic case is relevant in cosmology, where the charged primordial plasma screens the electric field, and the Universe's magnetic field is frozen for the magnetic properties to occur. 

The energy density and pressure for $E^2 = 0$ becomes
\begin{eqnarray} 
\rho & = & {F} f,  \nonumber \\
p & = & \frac{1}{3} {F} ~\left(f + 4 {F} f_{F}  \right),
\label{eq9}
\end{eqnarray}
where ${F} = \frac{1}{2} B^2$.

In \cite{hbenaoum2021}, Benaoum and {\"O}vg{\"u}n have proposed a function depending on two real parameters $\alpha$ and $\beta$ given by:
\begin{eqnarray}
f \left({F} \right) & = & \frac{{1}}{\left( \beta {F}^{\alpha} + 1 \right)^{1/\alpha}}, 
\label{eq10}
\end{eqnarray}
$\beta {F}^{\alpha}$ is dimensionless, $\beta$ is the nonlinearity parameter, and the usual Maxwell's electrodynamics Lagrangian is recovered when $\beta=0$. 

The energy density and pressure are:
\begin{eqnarray} 
\rho_B & = & \frac{{F}}{\left( \beta {F}^{\alpha} + 1 \right)^{1/\alpha}}, \nonumber \\
p_B & = & - \frac{{F}}{\left( \beta {F}^{\alpha} + 1 \right)^{1/\alpha}} + \frac{2}{3} \frac{B^2}{\left( \beta {F}^{\alpha} + 1 \right)^{1+ 1/\alpha}}.
\label{eq11}
\end{eqnarray}
The equation of state (EoS) satisfied by this NLED Lagrangian is: 
\begin{eqnarray}
p_B & = & \frac{1}{3} \rho_B \left(1 - 4 \beta \rho_B^{\alpha} \right), 
\label{eq12}
\end{eqnarray}
which clearly shows that when the non-linearly is turned off (i.e. $\beta=0$), it reduces to a radiation EoS.

In the context of the inflationary paradigm, we choose the background spacetime to be described by a homogeneous, isotropic and spatially flat metric, which takes the following form:
\begin{equation}
ds^{2}=-dt^{2}+a(t)^{2}\left[dr^{2}+r^{2}\left(d\theta^{2}+\sin^2\theta \,d\phi^{2}\right)\right],
\label{eq13}
\end{equation}
where $a(t)$ is the scale factor that governs spacetime evolution. 

For such a metric, the Friedmann equations can be easily computed, which results in, 
\begin{eqnarray}
H^2 & = & \left( \frac{\dot{a}}{a} \right)^2 = \frac{1}{3} ~\rho_B, \label{Friedmann} \\
3 \frac{\ddot a}{a} & = & - \frac{1}{2} \left( \rho_B + 3 p_B \right),
\label{eq14}
\end{eqnarray}
where $H = \dot{a}/a$ is the Hubble parameter.   

Using the conservation of the energy-momentum tensor $\nabla ^{\mu }T_{\mu \nu }=0$, the continuity equation  of the NLED is derived as,
\begin{eqnarray}
 \dot{\rho_B} + 3 H (\rho_B + p_B) =0.
\label{eq15}
\end{eqnarray}
The above equation can be readily integrated, yielding the following relation between the electromagnetic field strength ${F}$ and the scale factor $a$, 
\begin{eqnarray}
{F} & = & {F}_{\text{end}} \left(\frac{a_{\text{end}}}{a} \right)^4 = {F}_{0} \left(\frac{a_0}{a} \right)^4 = B^2/2.
\label{eq16}
\end{eqnarray}
It follows that:
\begin{eqnarray}
B & = & B_{\text{end}} ~\left(\frac{a_{\text{end}}}{a} \right)^2 = B_0 ~\left(\frac{a_0}{a} \right)^2,
\label{eq17}
\end{eqnarray}
where $a_{\text{end}}$ ($a_0$) is the value scale factor and ${F}_{\text{end}}= \frac{1}{2} B_{\text{end}}{^2}$ 
(${F}_{0}= \frac{1}{2} B_0{^2}$) is the value of the electromagnetic field at the end of inflation (at the current time) respectively.

Notice that in geometrized units, all the quantities have a dimension of the power of length $[L]$. In this system of units, a quantity which has $L^{n}T^{m}M^{p}$ in ordinary units converse to $L^{n+m+p}$. To recover nongeometrized units, we have to use the conversion factor $c^{m}(8 \pi G/c^2)^{p}$. Thus, the dimension of $B_{0}$ and $H_{0}$ is [$L^{-1}$] in geometrized units, and the conversion factors are $1 \text{Gauss}=1.44\times 10^{-24}\mathrm{cm}^{-1}$ and $H_{0}=h\,1.08\times 10^{-30}\mathrm{cm}^{-1}$, where  $h =(67.4\pm 0.5)\times 10^{-2}$ and $N_{\text{eff}}=2.99 \pm 0.17$ according to the Planck 2018 results \cite{Planck2018,riess}. Then, we are dealing with magnetic fields of the order $10^{-40}\mathrm{cm}^{-1} \lesssim B_0 \lesssim 10^{-33}\mathrm{cm}^{-1}$ in the present epoch. 
Then, we can obtain 
\begin{align}
{F}(z)&= \frac{1}{2}  B_0^2~\left( 1+z \right)^4= {F}_{\text{end}}~\left(\frac{1+z}{1+z_{\text{end}}}\right)^4,\\
{F}_{\text{end}} & = \frac{1}{2} B_{\text{end}}{^2}= \frac{1}{2}  B_0^2~\left(\frac{a_{\text{end}}}{a_0} \right)^{-4}= \frac{1}{2}  B_0^2~\left(1+z_{\text{end}}\right)^{4}
\end{align}
where we have introduced the redshift $z$, such that 
\begin{eqnarray}
1+z=\frac{a_0}{a}, \quad 1+z_{\text{end}}= \frac{a_0}{a_{\text{end}}},
\end{eqnarray}
where $a_0$ is the present value of the scale factor (we assume $a_0\neq 1$), $a_{\text{end}} = a (t_{\text{end}})$ is the scale factor evaluated at the end of inflation, and $z_{\text{end}}$ is the redshift at the end of inflation. 

Assuming that a Grand Unified Theory  (GUT) describes Nature, the grand unified epoch was the period in the evolution of the early Universe that followed the Planck epoch, beginning about $10^{-43}$ seconds after the Big Bang, in which the temperature of the Universe was comparable to the characteristic temperatures of the GUT. If the grand unification energy is taken as $10^{15}$ GeV, this corresponds to temperatures above $10^{27} K$. During this period, three out of four fundamental interactions, electromagnetism, the strong, and the weak, were unified into the electronuclear force. Gravity had separated from the electronuclear force at the end of the Planck era. Physical characteristics such as mass, charge, flavour, and colour were meaningless during the grand unification epoch. The GUT epoch ended at approximately $10^{-36}$ seconds after the Big Bang. At this point, several key events took place. The strong force separated from the other fundamental forces. Some parts of this decay process violated the conservation of baryon number and gave rise to a slight excess of matter over antimatter. It is also believed that this phase transition triggered the cosmic inflation process that dominated the Universe's evolution during the following inflationary epoch.
The inflationary epoch lasted from $10^{-36}$ seconds after the conjectured Big Bang singularity to some time values between $10^{-33 }$ and $10^{-32}$ seconds after the singularity. During the inflationary period, the Universe continued to expand, but at a slower rate. Then, we can take as characteristic values for $z_{\text{end}}$  at Grand Unified Theory (GUT) scale is $z_{\text{end}} \simeq z_{GUT} \simeq 10^{28}$,  or two orders of magnitude less, say $z_{\text{end}}\simeq 10^{-2} z_{GUT} \simeq 10^{26}$, or  $z_{\text{end}}\simeq  10^{10}$, i.e. two orders of magnitude above
nucleosynthesis. Hence, for the theoretical prior $z_{\text{end}}\simeq z_{GUT}$,   we have
$5 \times 10^{31} \mathrm{cm}^{-2}\lesssim {F}_{\text{end}} \lesssim 5 \times 10^{45} \mathrm{cm}^{-2} $.  For the theoretical prior $z_{\text{end}}\simeq  10^{10}$, we have $5 \times 10^{-41} \mathrm{cm}^{-2}\lesssim {F}_{\text{end}} \lesssim 5 \times 10^{-27} \mathrm{cm}^{-2}$. Taking a prior $z_{\text{end}}\simeq 10^{-2} z_{GUT} \simeq 10^{26}$ as an educated guess we have $5 \times 10^{23} \mathrm{cm}^{-2}\lesssim {F}_{\text{end}} \lesssim 5 \times 10^{37} \mathrm{cm}^{-2} $. 

Moreover,  the energy density and the pressure in terms of the scale factor $a$ can be expressed as, 
\begin{eqnarray}
\rho_B & = & \frac{\rho_0}{\left(1 + \left(\frac{a}{a_{\text{end}}} \right)^{4 \alpha} \right)^{1/\alpha}}, ~~~~~~ \; 
p_B = \rho_0 \frac{ -1 + \frac{1}{3} \left(\frac{a}{a_{\text{end}}} \right)^{4 \alpha}}{
\left(1 + \left(\frac{a}{a_{\text{end}}} \right)^{4 \alpha} \right)^{1+1/\alpha}},
\label{eq19}
\end{eqnarray}
where $\rho_0 = \rho_B (a =0)$ is the energy density at the early phase of the Universe. Plugging back  $\rho_B$ in equation \eqref{Friedmann}, it can be readily integrated, yielding the following solution for the Hubble parameter,
\begin{eqnarray}
t + const & = &  
\frac{1}{2 H} ~ {_2}F_1 \left(1,-\frac{1}{2 \alpha}; 1 - \frac{1}{2 \alpha}; \frac{1}{2} 
\left( \frac{H}{H_{\text{end}}} \right)^{2 \alpha}\right),
\label{eq38}
\end{eqnarray}
where $H_{\text{end}} = H (a_{\text{end}})$ is the value of the Hubble parameter at the end of the inflation.   

From the above equations, it follows, 
\begin{equation}
\lim_{a(t)\rightarrow 0}\rho_B (t)=\beta^{-1/\alpha},~~\lim_{a(t)\rightarrow
0} p_B (t)=-\beta^{-1/\alpha},  
\label{eq20}
\end{equation}%
\begin{equation}
~~\lim_{a(t)\rightarrow \infty }\rho_B (t)=\lim_{a(t)\rightarrow \infty
}p_B (t)=0. 
\label{eq21} 
\end{equation}%

It is easy to see that the Ricci scalar curvature, the Ricci tensor and the Kretschmann
scalar has no singularity at early/late stages, 
\begin{equation}
\lim_{a(t)\rightarrow 0}R(t)=4 \rho_0,  \quad 
\lim_{a(t)\rightarrow 0}R_{\mu \nu }R^{\mu \nu }=4 \rho_{0}^2, \quad
\lim_{a(t)\rightarrow 0}R_{\mu \nu \alpha \beta }R^{\mu \nu \alpha \beta }=\frac{8  \rho_{0}^2}{3}%
,  
\label{eq28}
\end{equation}%
\begin{equation}
\lim_{a(t)\rightarrow \infty }R(t)=\lim_{a(t)\rightarrow \infty }R_{\mu \nu
}R^{\mu \nu }=\lim_{a(t)\rightarrow \infty }R_{\mu \nu \alpha \beta }R^{\mu
\nu \alpha \beta }=0. 
\label{eq29}
\end{equation}%
The absence of singularities at early/late is an attractive feature peculiar to NLED.

\begin{figure}[hbtp]
\centering
  \begin{tabular}{@{}cc@{}}
  \includegraphics[width=.45\textwidth]{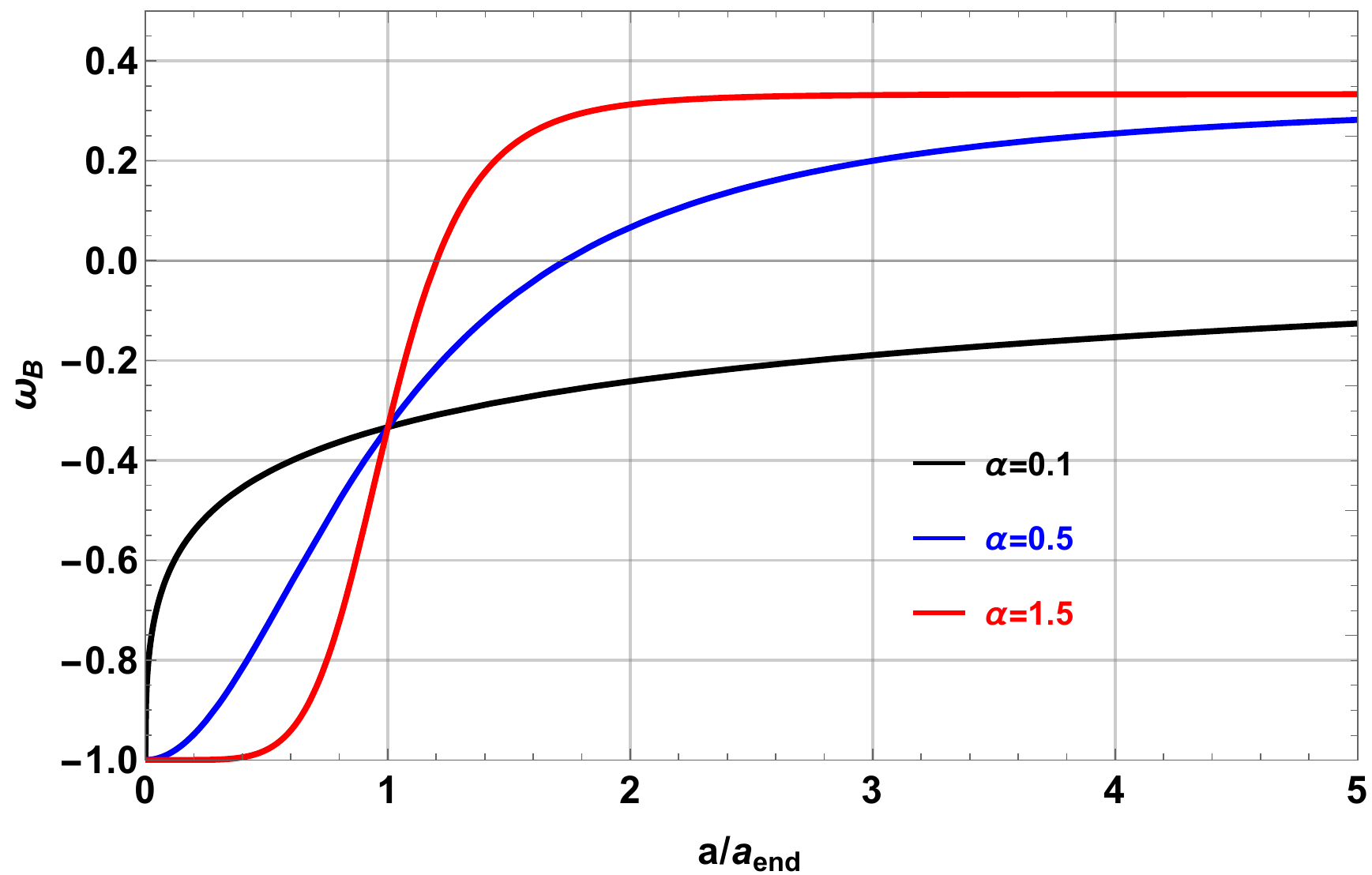} & \includegraphics[width=.45\textwidth]{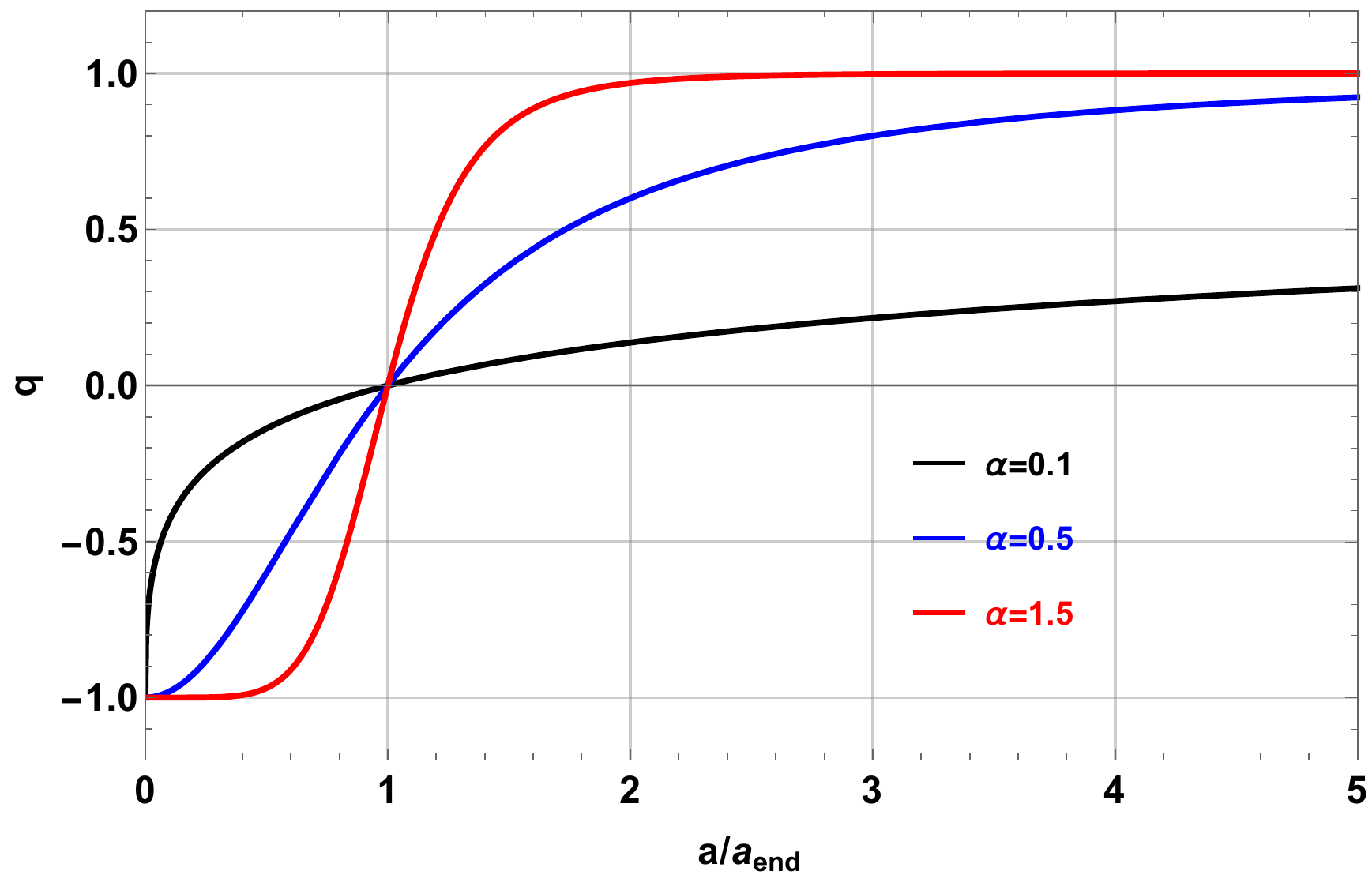} \\
\end{tabular}
\caption{ EoS of state parameter $\omega_B$ and deceleration $q$ as function of the scale factor $a$ for different values of $\alpha$.}
\label{fig:2}
\end{figure}

Now we concentrate on the evolution of the EoS parameter $\omega_B = p_B/\rho_B$, and the deceleration parameter, $q$, which are:
\begin{eqnarray}
\omega_B & = & \frac{p_B}{\rho_B} = 
\frac{ -1 + \frac{1}{3} \left(\frac{a}{a_{\text{end}}} \right)^{4 \alpha}}{1 + \left(\frac{a}{a_\text{end}} \right)^{4 \alpha} }, \quad 
q = \frac{1}{2} \left( 1 + 3 \omega_B \right) = 1 - 2 \beta \rho_B^{\alpha} .
\label{eq34} 
\end{eqnarray}
It follows that,  at small scale $ a \ll a_{\text{end}}$, 
\begin{eqnarray}
q & = & -1 +2  \left(\frac{a}{a_{\text{end}}} \right)^{4 \alpha},
\label{eq35}
\end{eqnarray} and at large scale $ a \gg a_{\text{end}}$, 
\begin{eqnarray}
q & = & 1 - 2  \left(\frac{a_{\text{end}}}{a} \right)^{4 \alpha}. 
\label{eq36}
\end{eqnarray}
Figure \ref{fig:2} shows the behaviour of the equation of state parameter $\omega_B$ and the deceleration $q$ as a function of the scale factor $a$ for different $\alpha=0.1,0.5,1.5$.

\begin{figure}[h]
    \centering
    \includegraphics[height=0.5\textwidth]{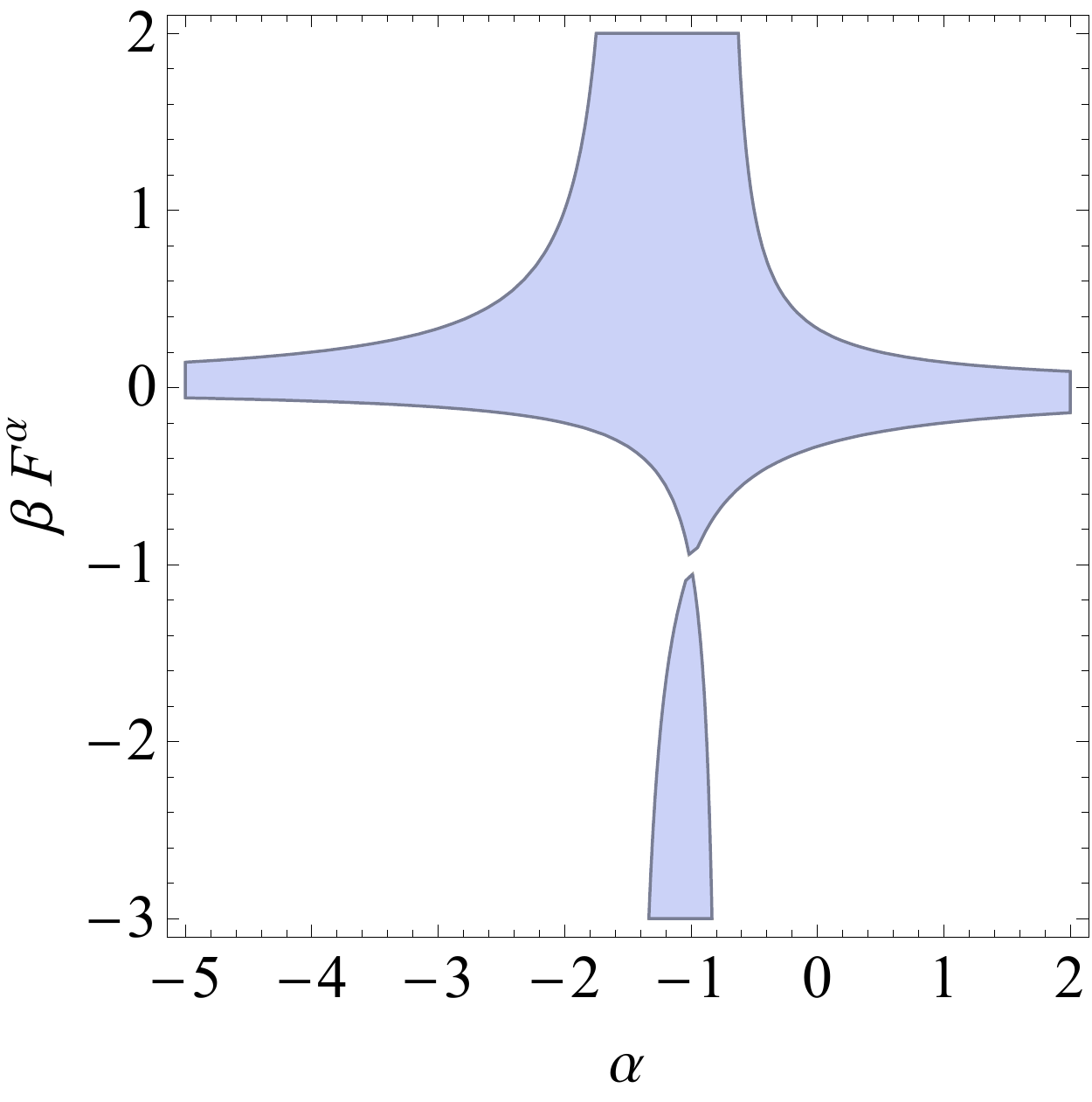}
    \caption{Regions of parameter space $\beta 
   {F}^{\alpha}$ vs $\alpha$ where a requirement of classical stability and causality is satisfied, i.e. $0 < c_s^2 \leq 1$.}
    \label{fig:1}
\end{figure}

The square speed of sound is
\begin{eqnarray}
c_s^2 & = & \frac{d p_B}{d \rho_B} = \frac{1-(4 \alpha +3) \beta  {F}^{\alpha }}{3 (\beta 
   {F}^{\alpha }+1)}.
\label{eq41}
\end{eqnarray}
Assuming $\beta 
   {F}^{\alpha}>0$, a requirement of classical stability and causality, i.e. $0 < c_s^2 \leq 1$, leads for
\begin{equation}
\left(\alpha <-\frac{3}{2}\land 0\leq \beta 
   {F}^{\alpha }<-\frac{1}{2 \alpha +3}\right)\lor
   \left(-\frac{3}{2}\leq \alpha \leq -\frac{3}{4}\land \beta
     {F}^{\alpha }\geq 0\right)\lor \left(\alpha
   >-\frac{3}{4}\land 0\leq \beta   {F}^{\alpha
   }<\frac{1}{4 \alpha +3}\right).
\label{eq42}
\end{equation} 
Now, we relax the condition $\beta \geq 0$, the region where classical stability and causality are required in Figure \ref{fig:1}.


\subsection{Cosmological Parameters}
In this section, we will demonstrate that it is indeed possible to have a proper inflationary phase in the early Universe described by NLED coupled with Einstein's gravity. To describe inflation, we use the 
e-folds number left to the end of inflation, 
\begin{eqnarray}
N = \ln \left( \frac{a_{\text{end}}}{a}\right).
\label{eq43}
\end{eqnarray}
Then, 
\begin{equation}
    \frac{d N}{d t}= -\frac{\dot{a}}{a}=-H. 
\end{equation}
In terms of e-folds number $N$, the energy density and the pressure read,
\begin{eqnarray}
\rho \left( N \right) & = & \frac{\rho_0}{\left( 1 + e^{- 4 \alpha N} \right)^{1/\alpha}} = 
\rho_{\text{end}} \left( 1 + \tanh \left(2 \alpha N \right) \right)^{1/\alpha}, \nonumber \\ 
p \left( N \right) & = & \rho \left( N \right) - f_1 \left( N \right),
\label{eq44}
\end{eqnarray}
where the function $f_1 \left( N \right)$ is given by: 
\begin{eqnarray*}
f_1 \left( N \right) = \frac{4}{3} \rho ~\left(1 - \beta \rho^{\alpha}  \right).
\end{eqnarray*}

The Hubble parameter will be:
\begin{eqnarray}
\rho \left( N \right) =  
H_{\text{end}} \left( 1 + \tanh \left(2 \alpha N \right) \right)^{\frac{1}{2 \alpha}}.
\label{eq45}
\end{eqnarray}

In this formalism, the slow-roll parameters are defined as:
\begin{eqnarray}
\epsilon & = & \frac{d \ln H}{d N} , \quad 
\eta = \epsilon + \frac{1}{2} \frac{d \ln \epsilon}{d N}.
\label{eq46}
\end{eqnarray}
Where the first slow-roll parameter $\epsilon$ relates to the acceleration measure during inflation, the second slow-roll parameter $\eta$ tells us how long the acceleration expression will be sustained.  

The slow-roll parameters become,
\begin{eqnarray}
\epsilon & = & 1 - \tanh \left( 2 \alpha N \right), \nonumber \\
\eta & = & 1 - \alpha - \left(1 + \alpha \right) ~\tanh \left( 2 \alpha N \right).
\label{eq47}
\end{eqnarray}
 Therefore the tensor-to-scalar $r$ and the scalar spectral index $n_s$ can be expressed as \cite{Bamba:2014wda}:
\begin{eqnarray}
r & = & 16 \epsilon = 16 \left(1 - \tanh \left( 2 \alpha N \right)\right), \nonumber \\
n_s & = & - 6 \epsilon + 2 \eta +1 = -3 - 2 \alpha + 2 \left(2 - \alpha \right) ~\tanh \left( 2 \alpha N \right) . 
\label{eq48} 
\end{eqnarray}

\begin{figure}[hbtp]
\centering
  \begin{tabular}{@{}cc@{}}
  \includegraphics[width=.45\textwidth]{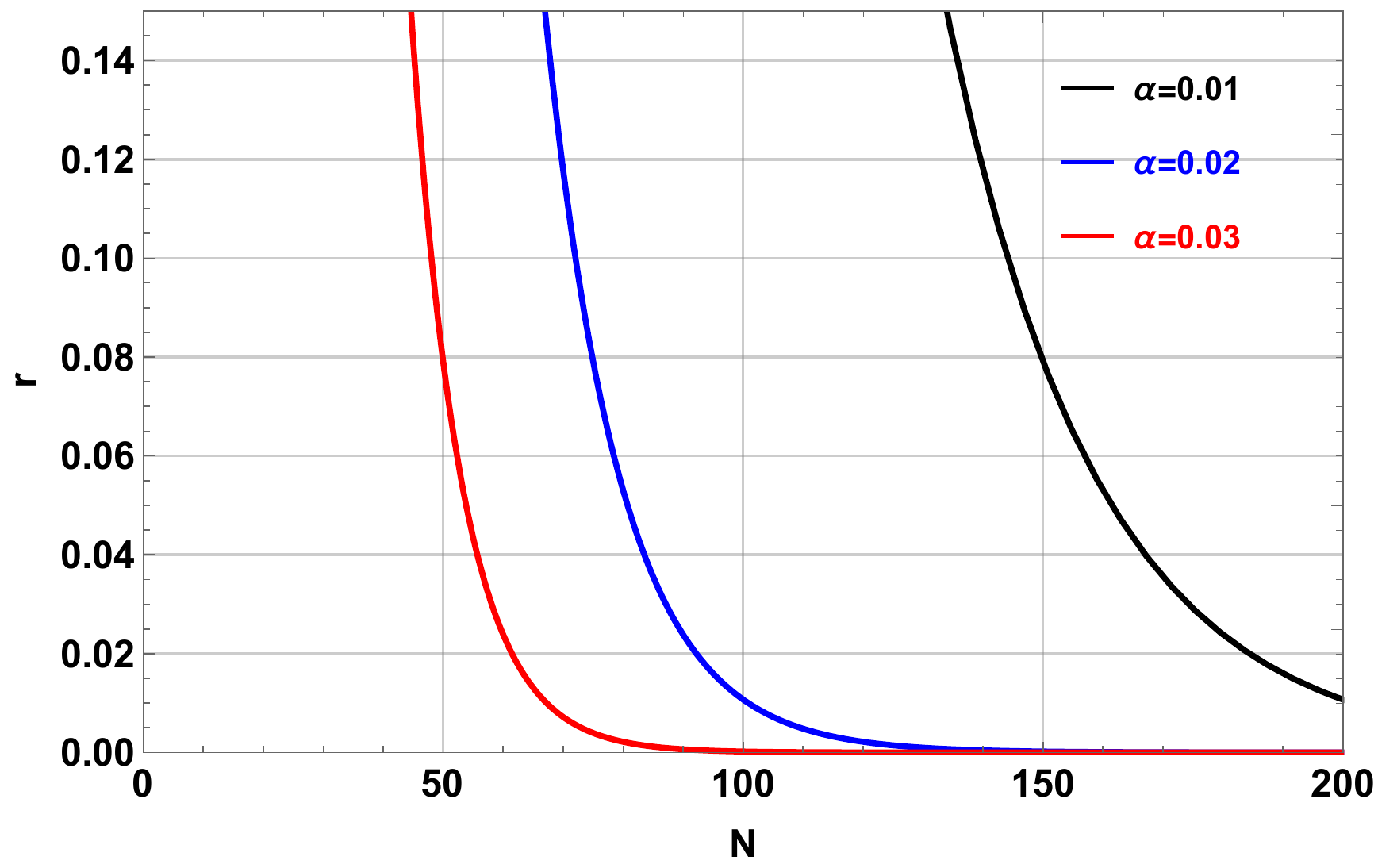} & \includegraphics[width=.45\textwidth]{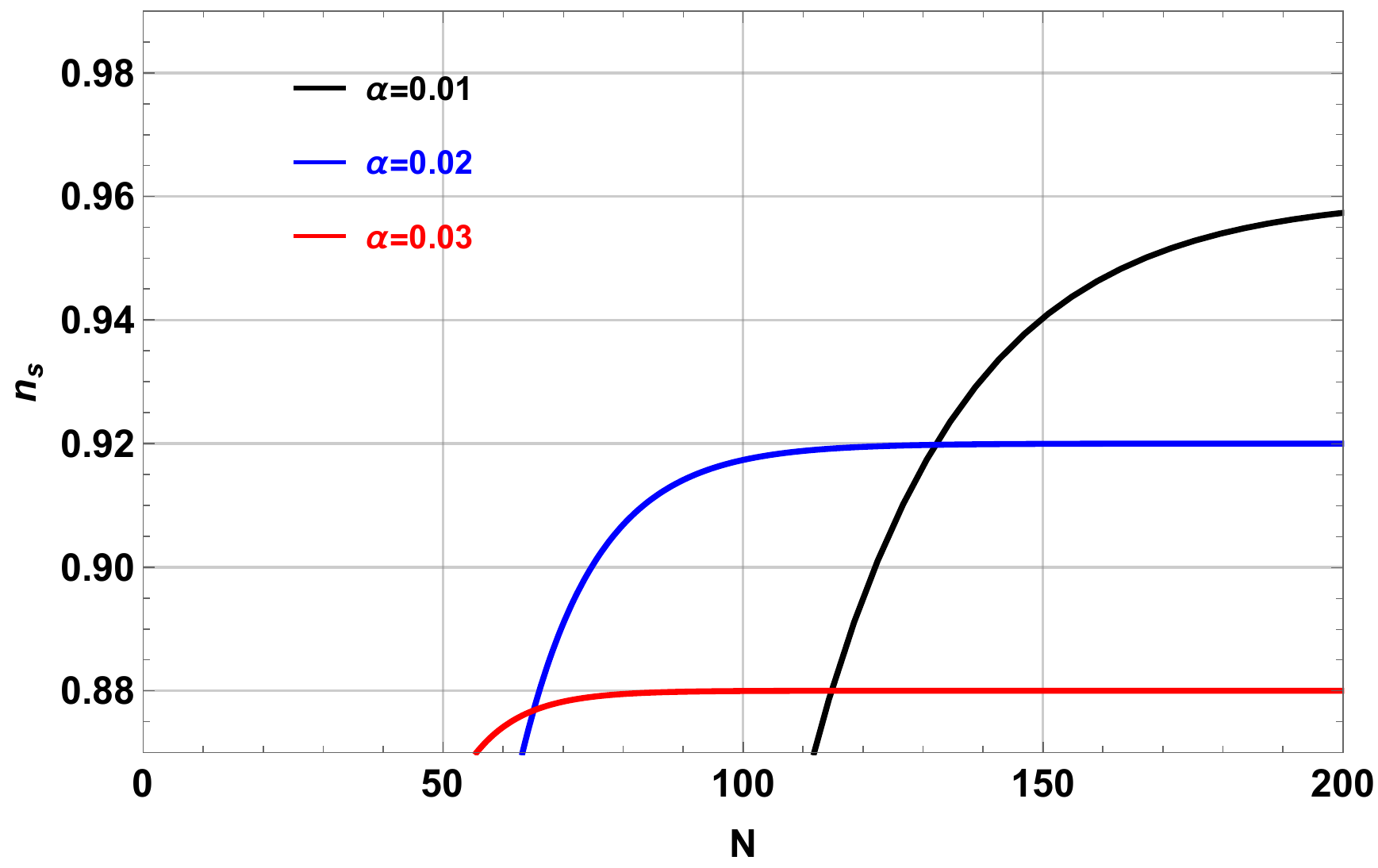} \\
\end{tabular}
\caption{ The tensor-to-scalar $r$ and the scalar spectral index $n_s$ as function of the e-folds number $N$ for different values of $\alpha$.}
\label{fig:3}
\end{figure}
In figure \ref{fig:3}, we plot the behaviour of $r$ and $n_s$ as a function of the number of e-fold for different values of $\alpha$.  
The Planck 2018 bounds on the spectral index and the tensor-to-scalar ratio are the following, 
\begin{eqnarray}
n_s = 0.9649 \pm 0.0042, ~~~~~~~ \; 
r < 0.064.
\label{eq49}
\end{eqnarray}
It can be seen from the figures that the Planck 2018 result rules out the model and that the e-folding number must be large and $\alpha$ extremely small to achieve successful inflation, away from the theoretical prior $50 < N < 60$.  

To overcome the major drawback of the model given by (\ref{eq10}), we propose the below general function 
$f \left( {F}\right)$ depending on three real parameters $A, \alpha$ and $\beta$ as an alternative to the function given by (\ref{eq10}),
\begin{eqnarray}
f \left( {F}\right) = \frac{{F}^{\frac{1}{4} (3 A -1)}}{\left(1 + \beta {F}^{\frac{3}{4}\alpha (A+1)}\right)^{1/\alpha}}.
\label{eq50}
\end{eqnarray}
It is clear that $A = \frac{1}{3}$ reproduces the NLED model given by (\ref{eq10}). The energy density of this model has the following e-folding number dependence, 
\begin{eqnarray}
\rho \left( N \right) = \rho_{\text{end}} ~\left( 1+ \tanh \left( \frac{3}{2} \alpha (A+1) N \right)\right)^{1/\alpha},
\label{eq51}
\end{eqnarray}
where $\rho_{\text{end}} = \frac{\rho_0}{2^{1/\alpha}} = \frac{1}{\left( 2 \beta \right)^{{1/\alpha}}}$. 

From the above equation, we obtain the Hubble parameter as,
\begin{eqnarray}
H \left( N \right) = H_{\text{end}}\left( 1+ \tanh \left( \frac{3}{2} \alpha (A+1) N \right)\right)^{1/2 \alpha}.
\label{eq52}
\end{eqnarray}
One can now get the scalar spectral index and the tensor-to-scalar as,
\begin{eqnarray}
r & = & 12 (A+1)\left( 1-  \tanh \left( \frac{3}{2} \alpha (A+1) N \right)\right),  \nonumber \\
n_s & = &  1- \frac{3}{2} (A+1) (2 + \alpha ) + \frac{3}{2} (A+1) (2 -\alpha) \tanh \left( \frac{3}{2} \alpha (A+1) N \right).
\label{eq53}
\end{eqnarray}

\begin{figure}[hbtp]
\centering
  \begin{tabular}{@{}cc@{}}
  \includegraphics[width=.45\textwidth]{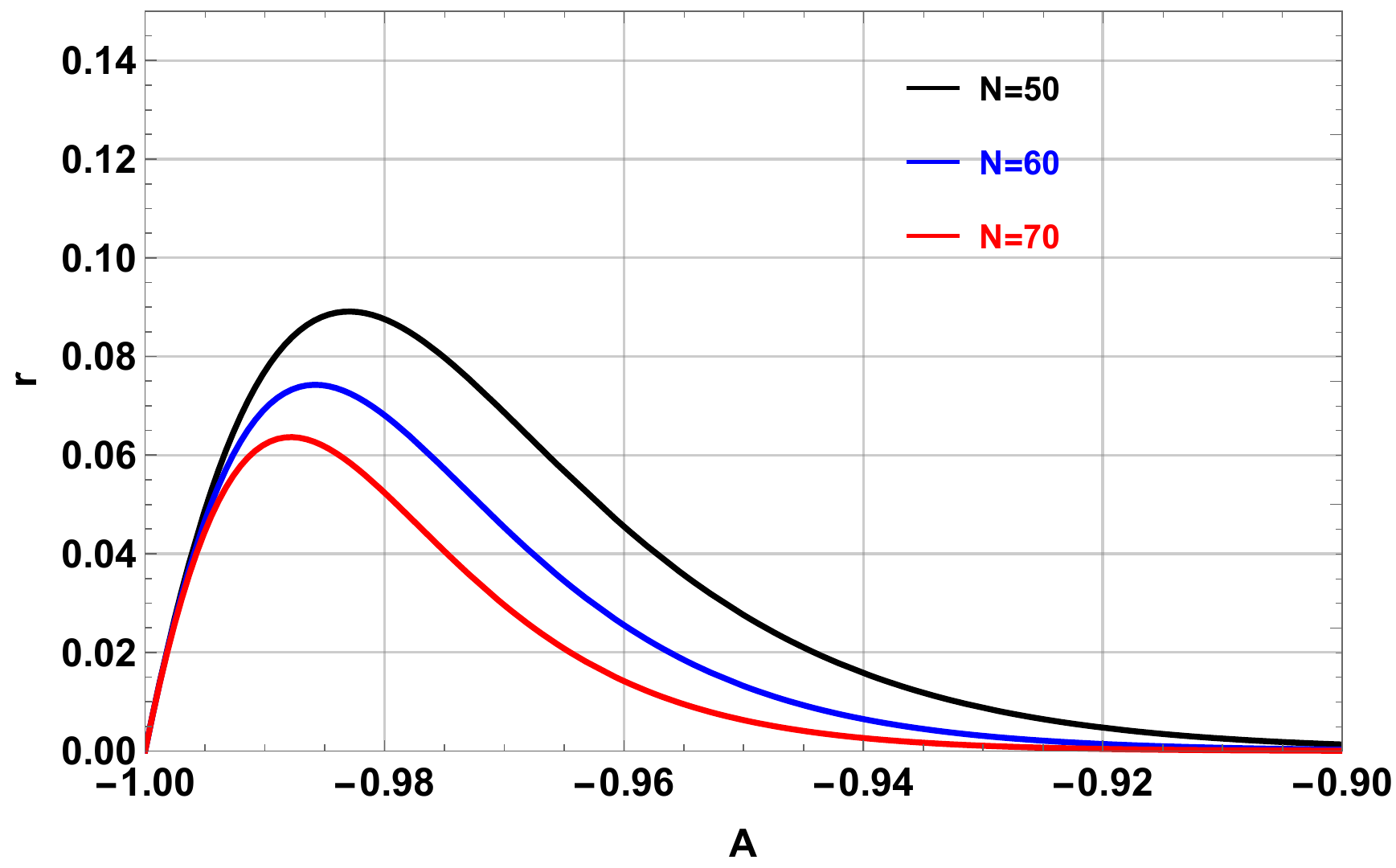} & \includegraphics[width=.45\textwidth]{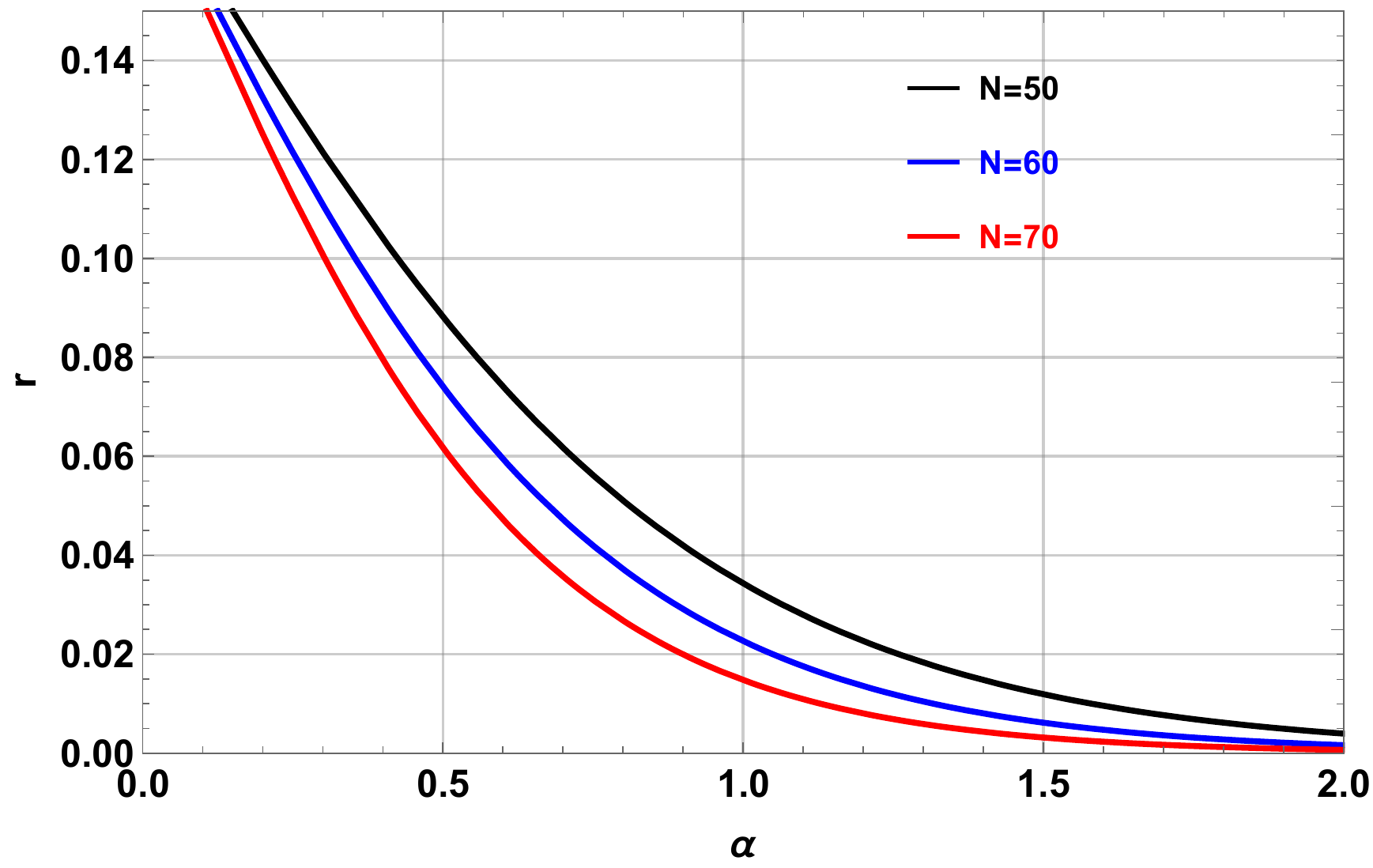} \\
\end{tabular} 
\caption{ The tensor-to-scalar ratio $r$ as function of the parameter $A$ with $\alpha=0.5$ and parameter $\alpha$ with $A=-0.985$ for $N=50,60,70$.}
\label{fig:4}
\end{figure}
Figure \ref{fig:4} displays the behavior of the tensor-to-scalar ratio $r$ versus the parameter $A$ by fixing 
$\alpha=0.5$ and the parameter $\alpha$ by fixing $A=-0.985$ for the e-folding number $N=50,60$ and $70$. 
We note from the figure that the value of $r$ increases till a maximum value and then decreases. We see that the bound $r< 0.064$ is achieved for $-1 < A < -0.995$ and $-0.967 < A < 0$. The figure also indicates that when $\alpha$ increases, the value of $r$ decreases and the bound $r < 0.064$ is satisfied for $\alpha >0.8$.

\begin{figure}[hbtp]
\centering
  \begin{tabular}{@{}cc@{}}
  \includegraphics[width=.45\textwidth]{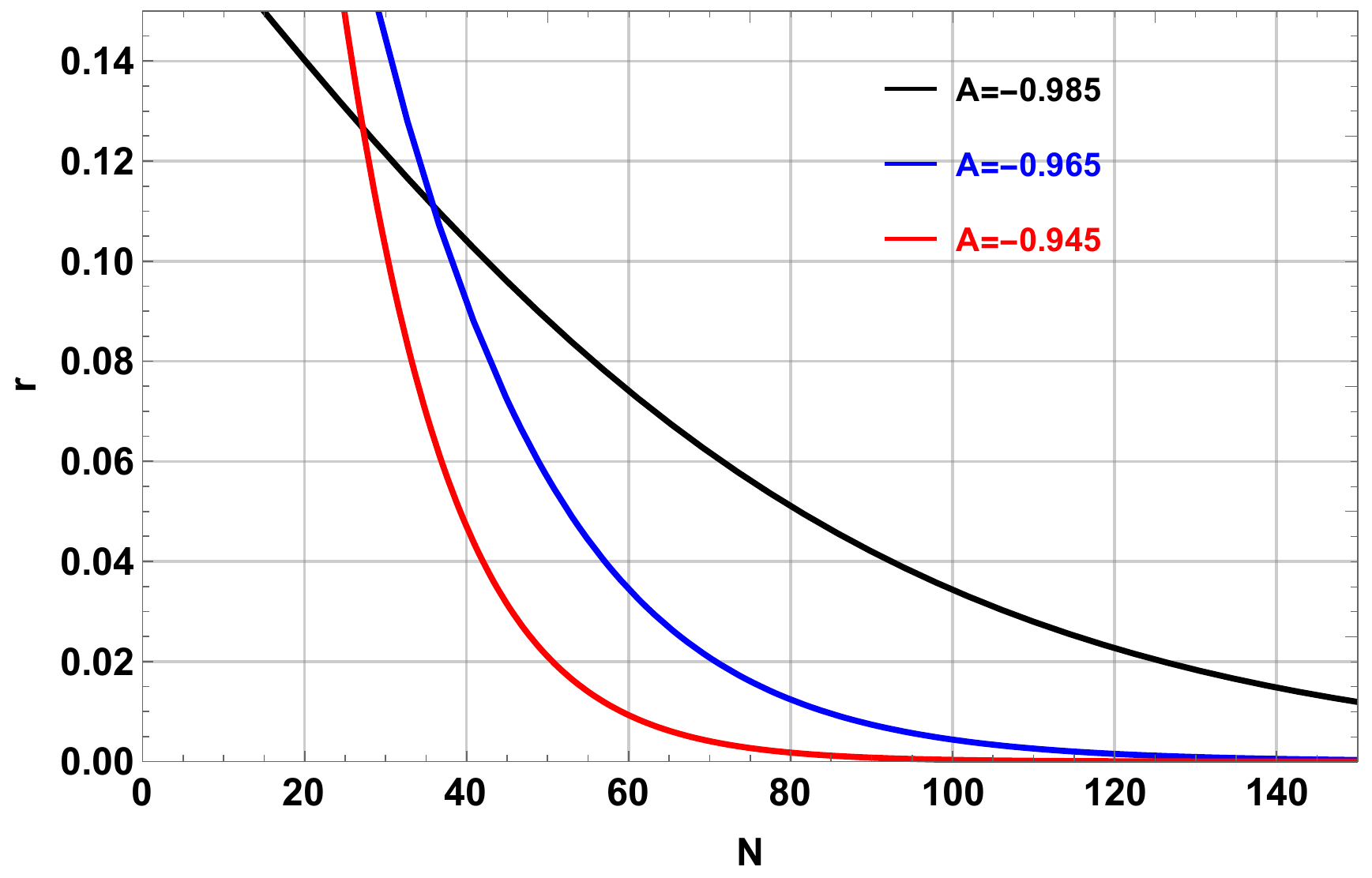} & \includegraphics[width=.45\textwidth]{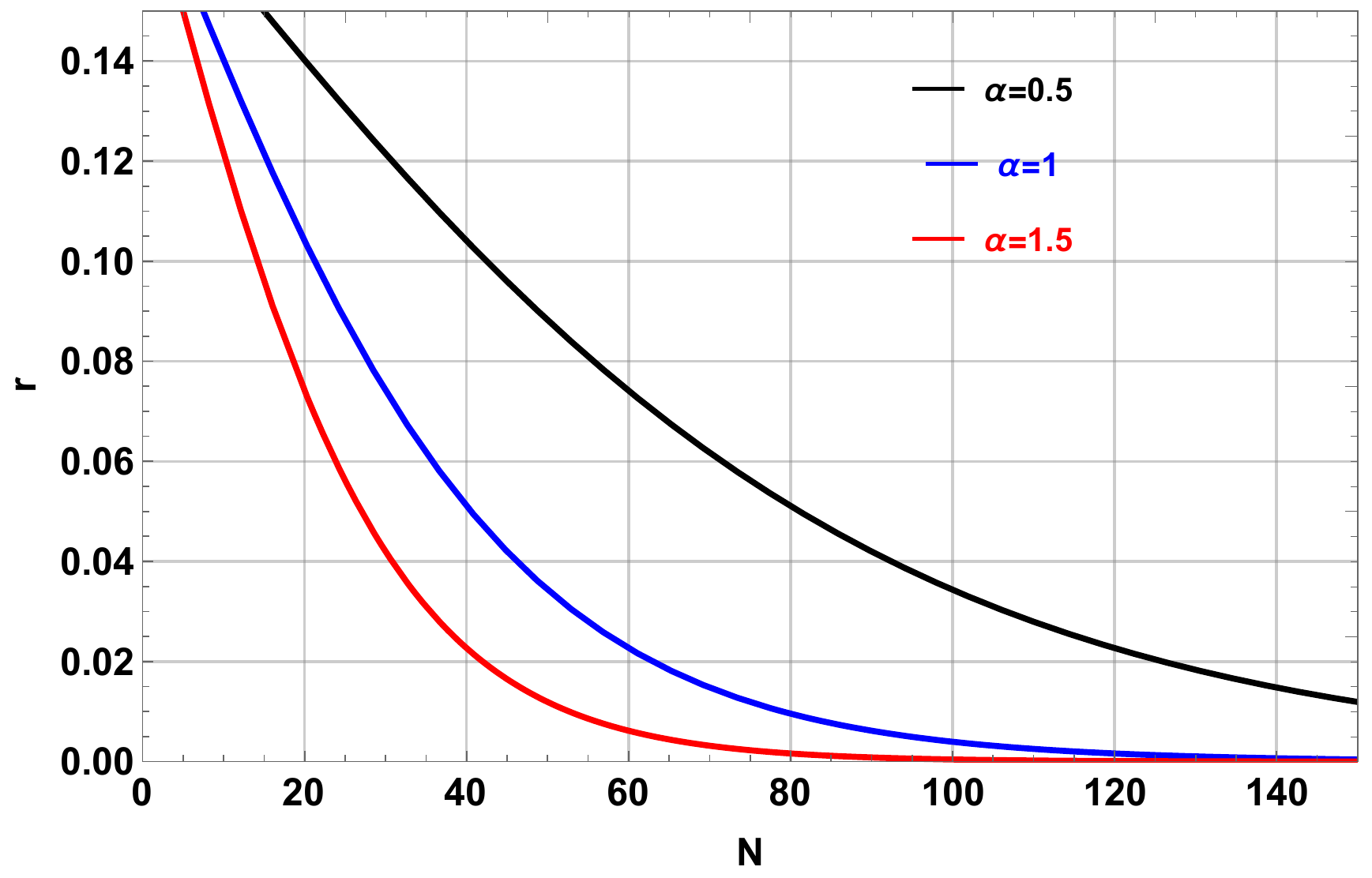} \\
\end{tabular} 
\caption{ The tensor-to-scalar ratio $r$ as function of the e-fold number $N$ for $\alpha =0.5$ with 
 $A=-0.985, -0,965, -0.945$ and for fixed $A=-0.985$ with $\alpha= 0.5, 1., 1.5$.}
\label{fig:5}
\end{figure}
In figure \ref{fig:5}, we plot the tensor-to-scalar ratio $r$ versus the e-folding number $N$ for fixed 
$\alpha =0.5$ with $A= -0.985,-0.965, -0.945$ and for fixed $A=-0.985$ with $\alpha = 0.5, 1., 1.5$. The main observation from the figure is that $r$ decreases with the e-folding number.  

\begin{figure}[hbtp]
\centering
  \begin{tabular}{@{}cc@{}}
  \includegraphics[width=.45\textwidth]{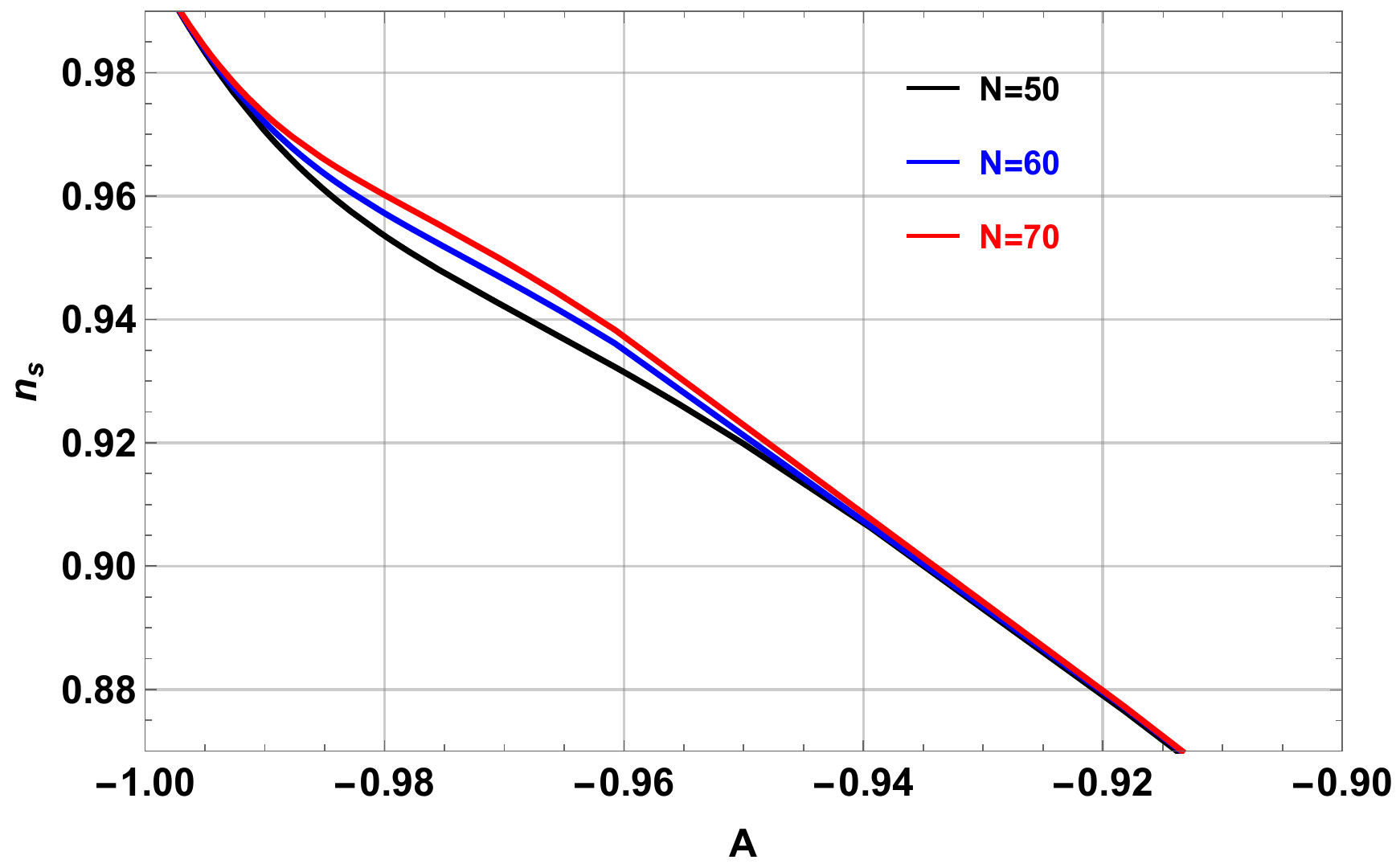} & \includegraphics[width=.45\textwidth]{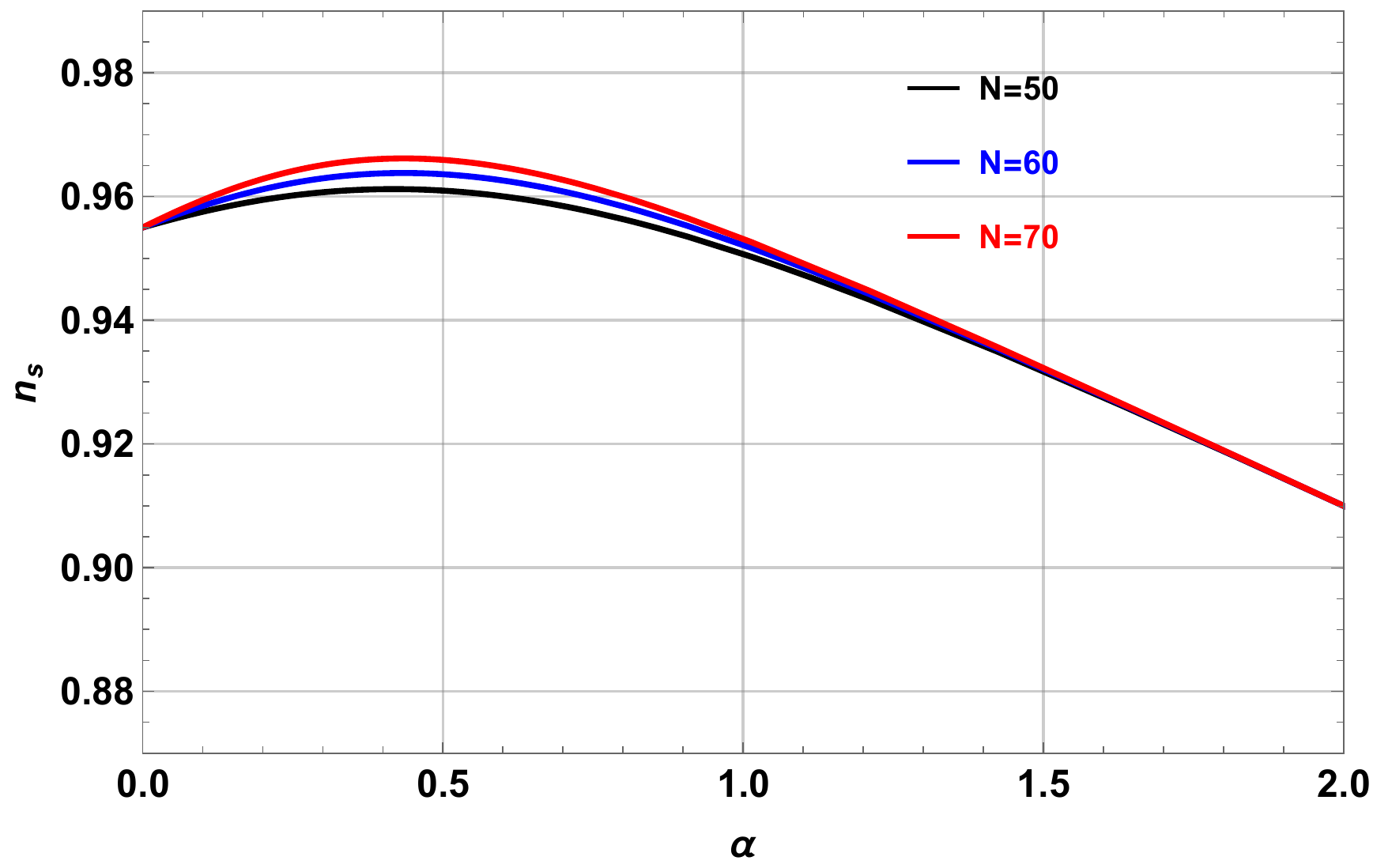} \\
\end{tabular} 
\caption{ The spectral index $n_s$ as function of the parameter $A$ with $\alpha=0.5$ and parameter $\alpha$ with $A=-0.985$ for $N=50,60,70$.}
\label{fig:6}
\end{figure}
Similarly, figure \ref{fig:6} shows the variation of the spectral index $n_s$ as a function of the parameter $A$ by fixing $\alpha=0.5$ and the parameter $\alpha$ by fixing $A=-0.985$ for the e-folding number $N$ equal to 50, 60 and 70. From the left figure in \ref{fig:6}, the spectral index $n_s$ decreases as $A$ increases. In the figure to the right, the spectral index $n_s$ increases to a maximum and decreases as $\alpha$ increases.  

\begin{figure}[hbtp]
\centering
  \begin{tabular}{@{}cc@{}}
  \includegraphics[width=.45\textwidth]{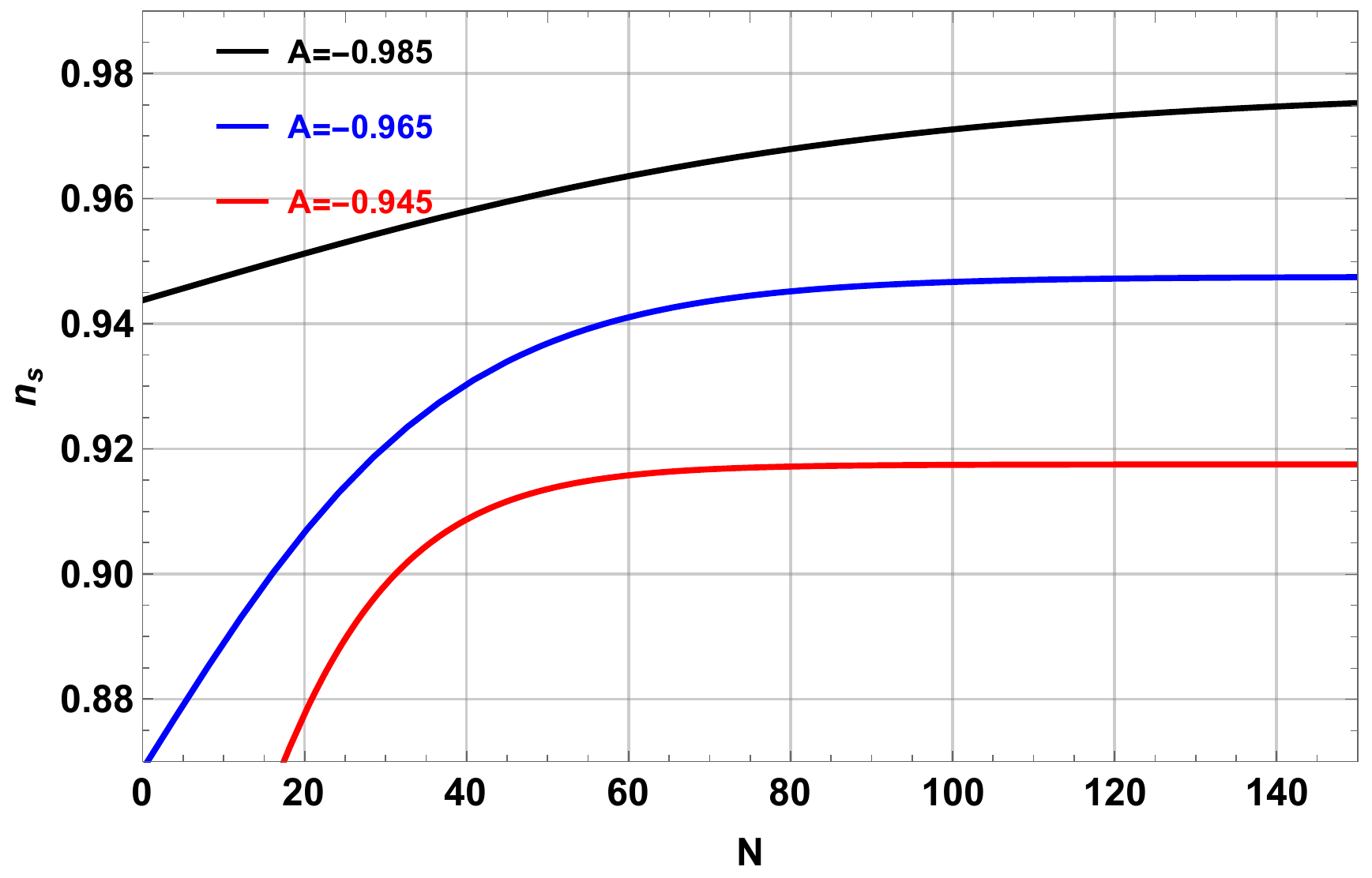} & \includegraphics[width=.45\textwidth]{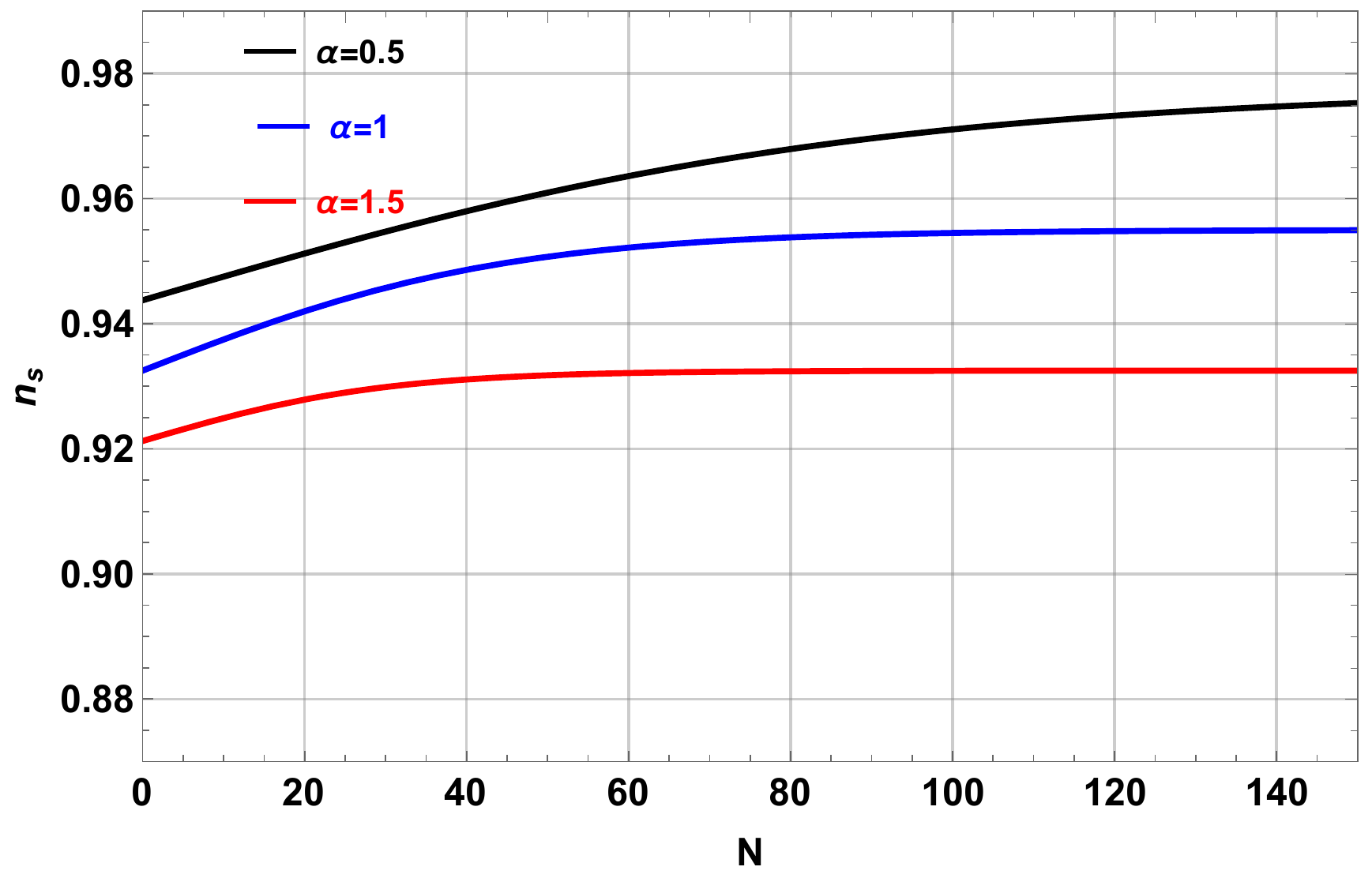} \\
\end{tabular} 
\caption{ The spectral index $n_s$ as function of the e-fold number $N$ for $\alpha =0.5$ with 
 $A=-0.985, -0,965, -0.945$ and for fixed $A=-0.985$ with $\alpha= 0.5, 1., 1.5$.}
\label{fig:7}
\end{figure}
\begin{figure}[hbtp]
\centering
  \begin{tabular}{@{}cc@{}}
  \includegraphics[width=.45\textwidth]{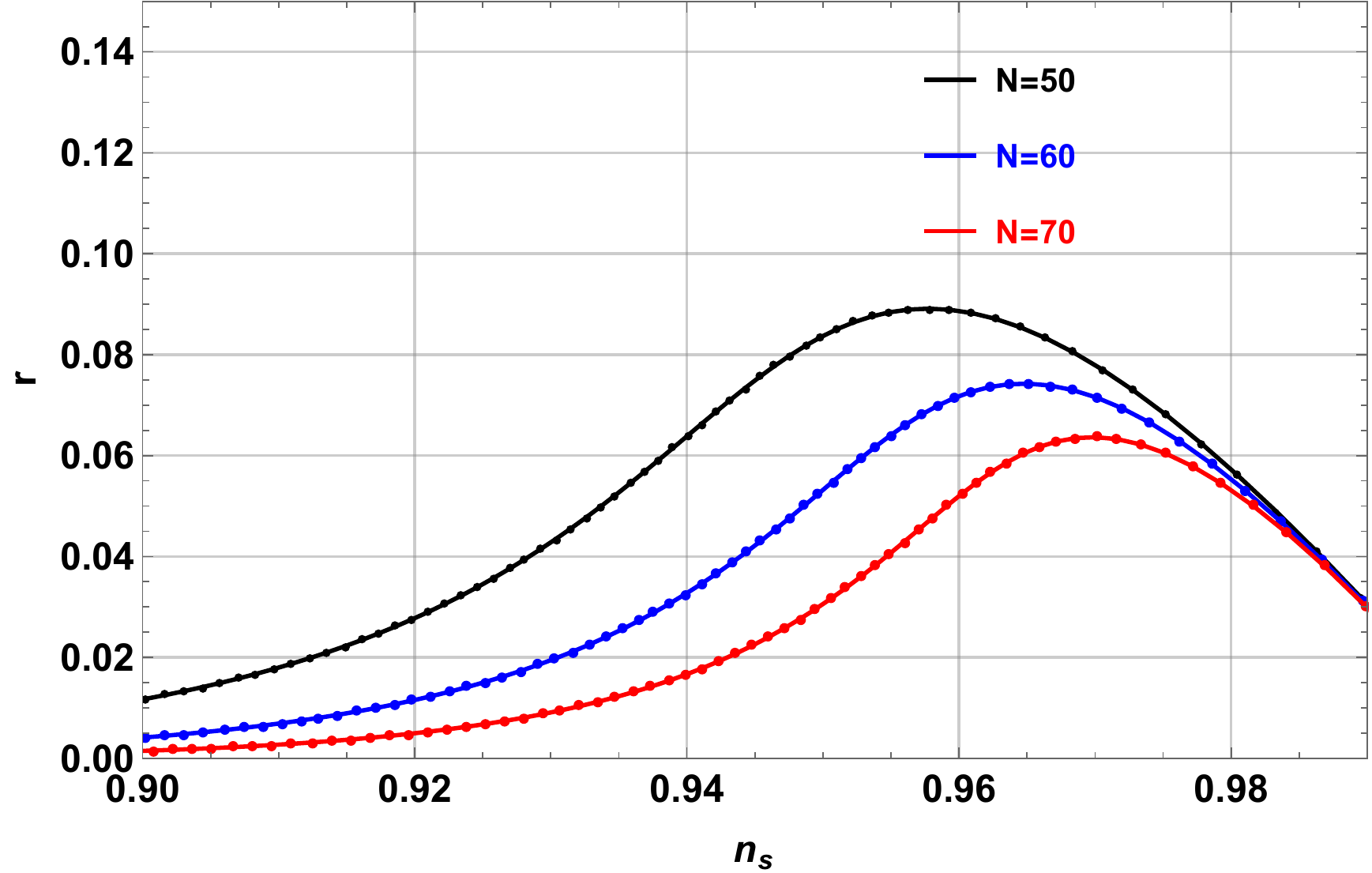} & \includegraphics[width=.45\textwidth]{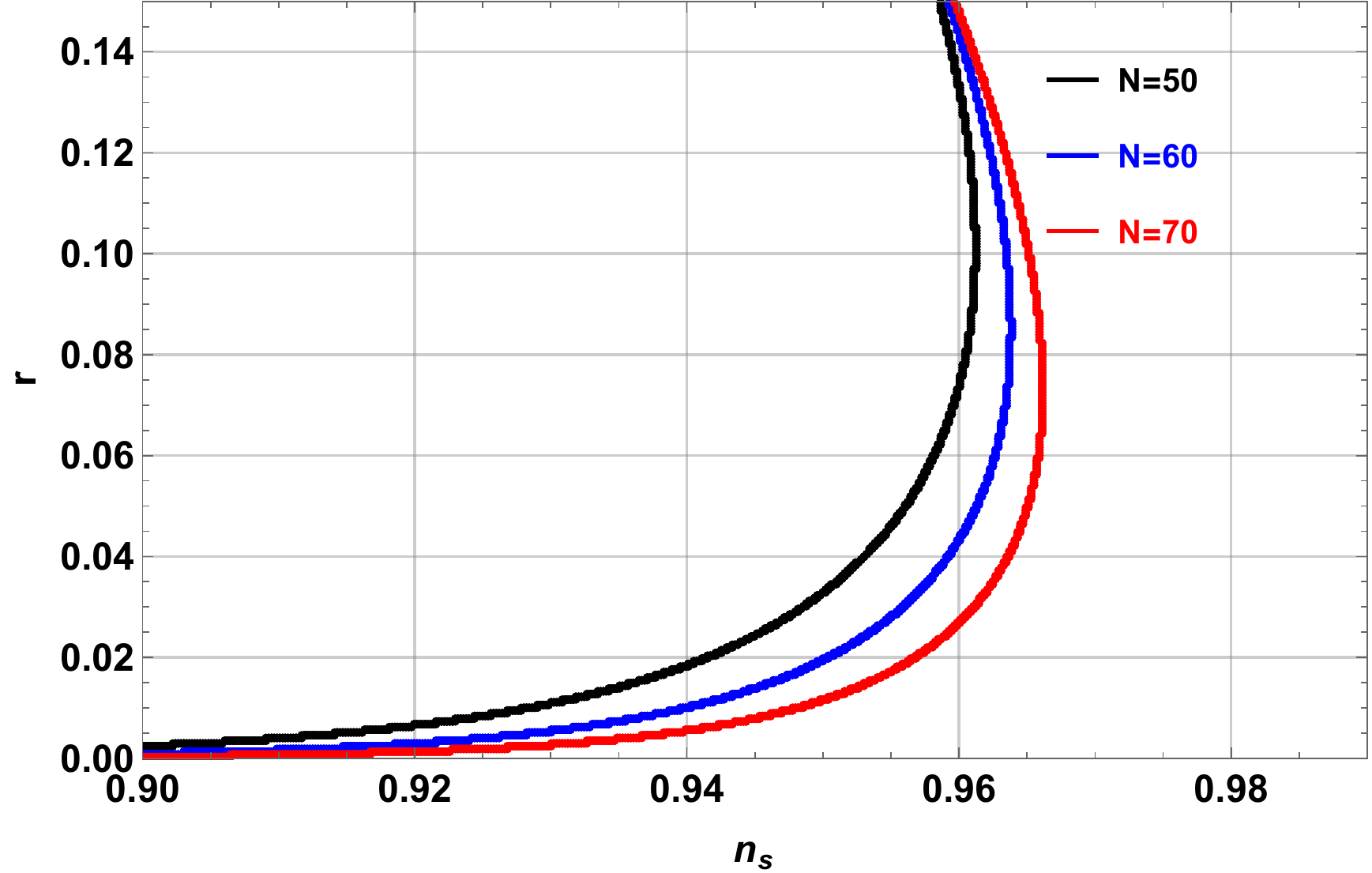} \\
\end{tabular} 
\caption{ The spectral index $n_s$ as function of the tensor-to-scalar ratio $r$ for $\alpha =0.5$ with 
 $A \in (-1,-0.9]$ and for fixed $A=-0.985$ with $\alpha \in (0,2]$.}
\label{fig:8}
\end{figure}

In figure \ref{fig:7}, we draw the spectral index $n_s$ as a function of the e-folding number $N$ for $\alpha =0.5$ with $A=-0.985, -0.965, -0.945$ and for $A=-0.985$ with $\alpha =0.5, 1., 1.5$. In the figure to the left, for fixed value $\alpha =0.5$, increasing $A$ leads to lower spectral index values $n_S$. A similar tendency in the figure to the right is observed by fixing $A=-0.985$ and increasing $\alpha$. Moreover, in order to present our results more transparently, we plot in figure \ref{fig:8}, 
$r$ versus $n_s$ by fixing $\alpha=0.5$ with $-1<  A  \leq -0.9$ and $A=-0.985$ with $0 <\alpha \leq 2$ for e-folding number $N$ equal to 50, 60 and 70. It shows that the current bounds on $n_s$ and $r$ are satisfied.   

\subsection{A route to a generalization}
\label{Sect. 2}

Interestingly, one may use the Lagrangian (\ref{eq2}) in such a way to describe all nonlinear electrodynamics 
models in the literature. While doing so, we want a unified prescription of all models. In the following, we provide an incomplete list of models that can be recovered from (\ref{eq2}).

\begin{table}[!h]  
\begin{center}
     \resizebox{\textwidth}{!}{ 
\begin{tabular}{|c|c|c|c|c|} 
\hline
\hline  
Model & Lagrangian ${\cal L}$    & $f \left( {F}\right)$ & Maxwell's limit & References \\ 
\hline
Born-Infeld model & $- \alpha^2 \left( \sqrt{1+ \frac{2 {F}}{\alpha^2}} - 1\right)$ & 
$\frac{\alpha^2}{{F}} \left( \sqrt{1+ \frac{2 {F}}{\alpha^2}} - 1\right)$ & 
$\alpha \gg \sqrt{2 |{F}|}$ & \cite{BIa,BIb,BIc} \\
\hline
De Lorenci et al. model & $- {F} + 16 \alpha {F}^2$ & $1- 16 \alpha {F}$ & $\alpha \rightarrow 0$ &  
\cite{DeLorenci} \\
\hline
Novello's Toy model & $- {F} + 16 \alpha^2 {F}^2 - \frac{\beta}{{F}}$ & $1 - 16 \alpha {F} + \frac{\beta}{{F}^2}$ & $\alpha \rightarrow 0, \beta \rightarrow 0$ & \cite{novello6} \\
\hline
Kruglov's model A & $- {F} \left(1 - \frac{\alpha}{2 \beta {F} +1} \right)$ & $1 + \frac{1}{2 \beta {F}}$ & $\alpha \rightarrow 0$ &  \cite{kruglov4}\\ 
\hline 
Kruglov's model B & $- \frac{F}{\beta {F} +1}$ & $\frac{1}{\beta {F} +1}$ & $\beta \rightarrow 0$ &  \cite{kruglov4}\\ 
\hline
Kruglov's model C & $-{F} \text{sech}^2\left(\sqrt[4]{\left| {F} \beta \right| }\right)$ & $\text{sech}^2\left(\sqrt[4]{\left| {F} \beta \right| }\right)$ & $\beta {F} \rightarrow 0$ & \cite{kruglov6}\\
\hline
{\"O}vg{\"u}n's exponential correction model & $\frac{- {F} e^{- \alpha {F}}}{\alpha {F} + \beta}$ & 
 $\frac{e^{- \alpha {F}}}{\alpha {F} + \beta}$ & $\alpha \rightarrow 0, \beta \rightarrow 1$ & \cite{Ovgun:2017iwg} \\
\hline
Benaoum and {\"O}vg{\"u}n model & $\frac{- {F}}{(\beta {F}^{\alpha}+ 1)^{1/\alpha}}$ & 
$\frac{1}{(\beta {F}^{\alpha}+ 1)^{1/\alpha}}$ & $\beta \rightarrow 0$ &  \cite{hbenaoum2021} \\
\hline
\end{tabular}}
\caption{\label{tab:Table 1} Incomplete List of some NLED models.}
\end{center}
\end{table}  
Initially, the idea is to consider that $f = f \left( {F} \right)$ is an arbitrary function. Then we proceed to specify some suitable conditions the model has to satisfy. From equation 
(\ref{eq9}), the energy density, the pressure and the equation of state parameter for purely magnetic field 
(i.e. $\vec{E} = \vec{0}$ and ${F} = \frac{1}{2} B^2$), we have 
\begin{eqnarray}
\rho_B & = & {F} f,  ~~~~~~~ 
p_B = \frac{1}{3} {F} \left( f + 4 {F} f_{{F}} \right),  ~~~~~~~
\omega_B = \frac{p_B}{\rho_B} = \frac{1}{3} + \frac{4}{3} {F} \frac{f_{{F}}}{f},
\label{eq54}
\end{eqnarray}
where we use the notations $f_{{F}}= f'({F}), f_{{F} {F}}= f''({F}), \ldots$. 
The Ricci scalar, which represents the curvature of the spacetime, is calculated by using Einstein field equations and the energy-momentum tensor, 
\begin{eqnarray}
R = \rho_B - 3 p_B = - 4 {F}^2 f_{{F}}.
\label{eq55}
\end{eqnarray}
The Ricci tensor squared $R^{\mu \nu} R_{\mu \nu}$ and the Kretschmann $R^{\mu \nu \alpha \beta} R_{\mu \nu \alpha \beta}$ are also obtained, 
\begin{eqnarray}
R^{\mu \nu} R_{\mu \nu} = \rho_B^2 + 3 p_B^2 = \frac{1}{3} {F}^2 f^2 \left( 4 + 
2 {F} \frac{f_{{F}}}{f} + {F}^2 \frac{f_{{F}}^2}{f^2} \right), \nonumber \\
R^{\mu \nu \alpha \beta} R_{\mu \nu \alpha \beta} = \frac{8}{3} {F}^2 f^2 \left(1 + 
2 {F} \frac{f_{{F}}}{f} + 2 {F}^2 \frac{f_{{F}}^2}{f^2} \right).
\label{eq56}
\end{eqnarray}
The squared sound of speed is
\begin{eqnarray}
c_s^2 = \frac{d p_B}{d \rho_B} = \frac{1}{3} + \frac{4}{3} \frac{2 {F} f_{{F}} + {F}^2 f_{{F} {F}}}{f + {F} f_{{F}}}.
\label{eq57}
\end{eqnarray}

In the following, we define the conditions that should satisfy a viable NL$f$ED: 
\begin{enumerate}
\item Removal of singularities at early/late phase of the Universe. In flat spacetime, a sufficient condition for the Ricci scalar, the Ricci tensor squared, and the Kretschmann scalar to not be singular is to choose $f = f \left( {F} \right)$ such that the energy density and the pressure are finite in the limit of large ${F}$ and small ${F}$. Interestingly, in our approach, the pressure can be expressed in terms of the energy density and its derivative as:
\begin{eqnarray}
p_B = \rho_B \left(-1 +\frac{4}{3} \frac{d \rho_B}{d \ln {F}} \right).
\label{eq58}
\end{eqnarray}
That implies that in order to remove singularities to the early/late phase of the Universe, both the energy $\rho_B$ density and its derivative $\frac{d \rho_B}{d \ln {F}}$ have to be finite at $a \rightarrow 0$ (large ${F}$) and $a \rightarrow \infty$ (small ${F}$).  

\item Early/late time radiation/dark energy domination phase for large/small magnetic field: 
\begin{equation}
\lim_{{F} \to \infty /
{F} \to 0} \frac{d \rho_B}{d \ln {F}} = 1, 0  \implies \lim_{a\rightarrow 0 /
a \rightarrow \infty}\omega_B =\frac{1}{3}, -1.
\label{eq59}
\end{equation}

\item Condition of the accelerated Universe $\rho_{B}+3p_{B}<0$ with the sources of NL$f$ED fields are,
\begin{eqnarray}
\rho_{B}+3 p_{B} & = & {F} f + {F} \left( f + 4 {F} f_{{F}} \right)  \nonumber \\ 
& = & 2 \rho_B \left( - 1 + \frac{2}{3} \frac{d \ln \rho_B}{d \ln {F}} \right).
\label{eq60}
\end{eqnarray}
It gives,
\begin{eqnarray}
\frac{ \ln \rho_B}{d \ln {F}} < \frac{3}{2}.
\label{eq61}
\end{eqnarray}
This inequality has to be satisfied for large $B$, that is, acceleration during the inflationary phase, and for small $B$, which corresponds to late-time acceleration.  
\item Classical stability: 
\begin{enumerate}
\item Causality of the Universe: the speed of the sound should be lower than the local light speed ($c_{s}<1$) \cite{sound}.

\item To avoid the Laplacian instability, we require the conditions that must be positive ($c_{s}^{2}>0$).
\end{enumerate} 

Classical stability and causality give,
\begin{equation}
0< \frac{1}{3} + \frac{4}{3} \frac{2 {F} f_{{F}} + {F}^2 f_{{F} {F}}}{f + {F} f_{{F}}}<1,
\label{eq62}
\end{equation}
\end{enumerate}
which implies, 
\begin{eqnarray}
- \frac{1}{4} < \frac{d}{d \rho_B} \left( {F}^2 f_{{F}} \right) < \frac{1}{2} .
\label{eq63}
\end{eqnarray}

\subsection{Integrability and connection with the observables}
\label{Sect:obs_2}
 
In this section, we comment on the integrability of the system at hand. Moreover, we calculate some observables in the e-folding number $N$.  

From the first equation of Friedmann's equations (\ref{eq14}), one can obtain  an equation which shows the conservation of energy for a particle moving in an effective potential $V_{\text{eff}} \left( a \right)$:
\begin{equation}
\frac{1}{2}\dot{a}^{2}+V_{\text{eff}}(a)=0, 
\label{eq64}
\end{equation}
with 
\begin{eqnarray} 
\quad V_{\text{eff}}(a) = - \frac{1}{6} a^2 {F} f.
	\label{eq65}
\end{eqnarray}
For a positive scale factor, we get
   \begin{equation}
  \dot a=\frac{a}{\sqrt{3}} \sqrt{{F} f},
	\label{eq66}
   \end{equation}
  which can be solved by quadratures:
  \begin{equation}
  \frac{t}{\sqrt{3}}  =\int \frac{1}{\sqrt{{F} f}} \, dN.
	\label{eq66b}
  \end{equation}
For the general function given by (\ref{eq50}), we obtain,
\begin{equation}
 t =   - \frac{2}{(A+1) \sqrt{3 \rho_B}}  ~_2F_1\left(-\frac{1}{2 \alpha },-\frac{1}{2 \alpha };1-\frac{1}{2 \alpha };\frac{1}{2} \left( \frac{\rho_B}{\rho_{\text{end}}} \right)^{\alpha} \right) + C.
\end{equation}
As a function of $N:= \ln (a_{\text{end}}/a)$, the magnetic field $B \left(N \right)$, the magnetic field strength ${F} \left(N \right)$, the Hubble parameter $H \left( N \right)$ and the deceleration $q \left( N \right)$ are (recall $N=0$ at the end of inflation):
\begin{eqnarray}
B \left(N \right) & = & B_{\text{end}} ~e^{2 N}, \nonumber \\
{F} \left(N \right) & = & {F}_{\text{end}} ~e^{4 N}, \nonumber \\
H \left(N \right) & = & \sqrt{\frac{1}{3}{F}_{\text{end}} ~e^{4 N}  f({F}_{\text{end}} ~e^{4 N} )} =  H_{\text{end}} ~e^{2 N} ~\sqrt{\frac{f \left({F}_{\text{end}}~e^{4 N} \right) }{f \left({F}_{\text{end}}\right)}}, \nonumber \\
q \left(N \right) & = & - 1- \frac{\dot{H}}{H^2} = -1 + \frac{d \ln H}{d N} =1 + \frac{2 {F}_{\text{end}} e^{4 N} f'\left({F}_{\text{end}} e^{4
   N}\right)}{f\left({F}_{\text{end}} e^{4 N}\right)},
\label{eq67}
\end{eqnarray}
where 
\begin{equation}
    H_{\text{end}}= \sqrt{\frac{1}{3}{F}_{\text{end}}f({F}_{\text{end}})}.
\end{equation}
Recall that at large scale $ a \gg a_{\text{end}}$, $N<0$. 
The relation between $N$ and the redshift is 
\begin{equation}
    N= \ln \left[\frac{1+z}{1+z_{\text{end}}}\right]. \label{eqz55}
\end{equation}
Then, we have
\begin{eqnarray}
B \left(z\right) & = & B_{\text{end}} ~\left[\frac{1+z}{1+z_{\text{end}}}\right]^2, \nonumber \\
{F} \left(z\right) & = & {F}_{\text{end}} ~\left[\frac{1+z}{1+z_{\text{end}}}\right]^4, \nonumber \\
H \left(z\right) & = &   H_{\text{end}} ~\left[\frac{1+z}{1+z_{\text{end}}}\right]^2 ~\sqrt{\frac{f \left({F}_{\text{end}}~\left[\frac{1+z}{1+z_{\text{end}}}\right]^4 \right) }{f \left({F}_{\text{end}}\right)}}, \nonumber \\
q \left(z\right) & = & 1 + \frac{2 {F}_{\text{end}} \left[\frac{1+z}{1+z_{\text{end}}}\right]^4 f'\left({F}_{\text{end}} \left[\frac{1+z}{1+z_{\text{end}}}\right]^4\right)}{f\left({F}_{\text{end}} \left[\frac{1+z}{1+z_{\text{end}}}\right]^4\right)}.
\label{zeq67}
\end{eqnarray}

\begin{figure}[hbtp]
\centering
  \begin{tabular}{@{}cc@{}}
  \includegraphics[width=0.45\textwidth]{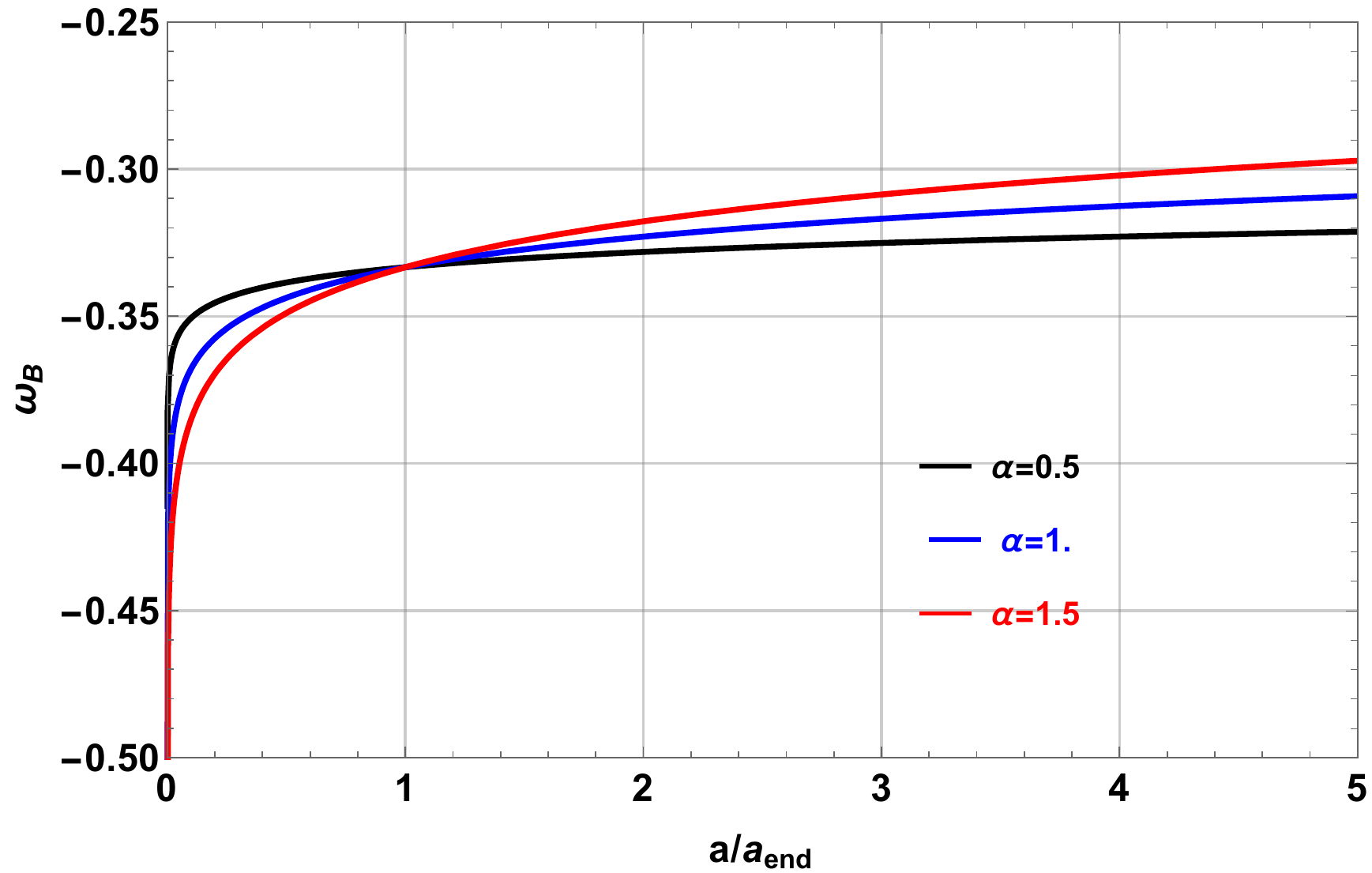} & \includegraphics[width=0.45\textwidth]{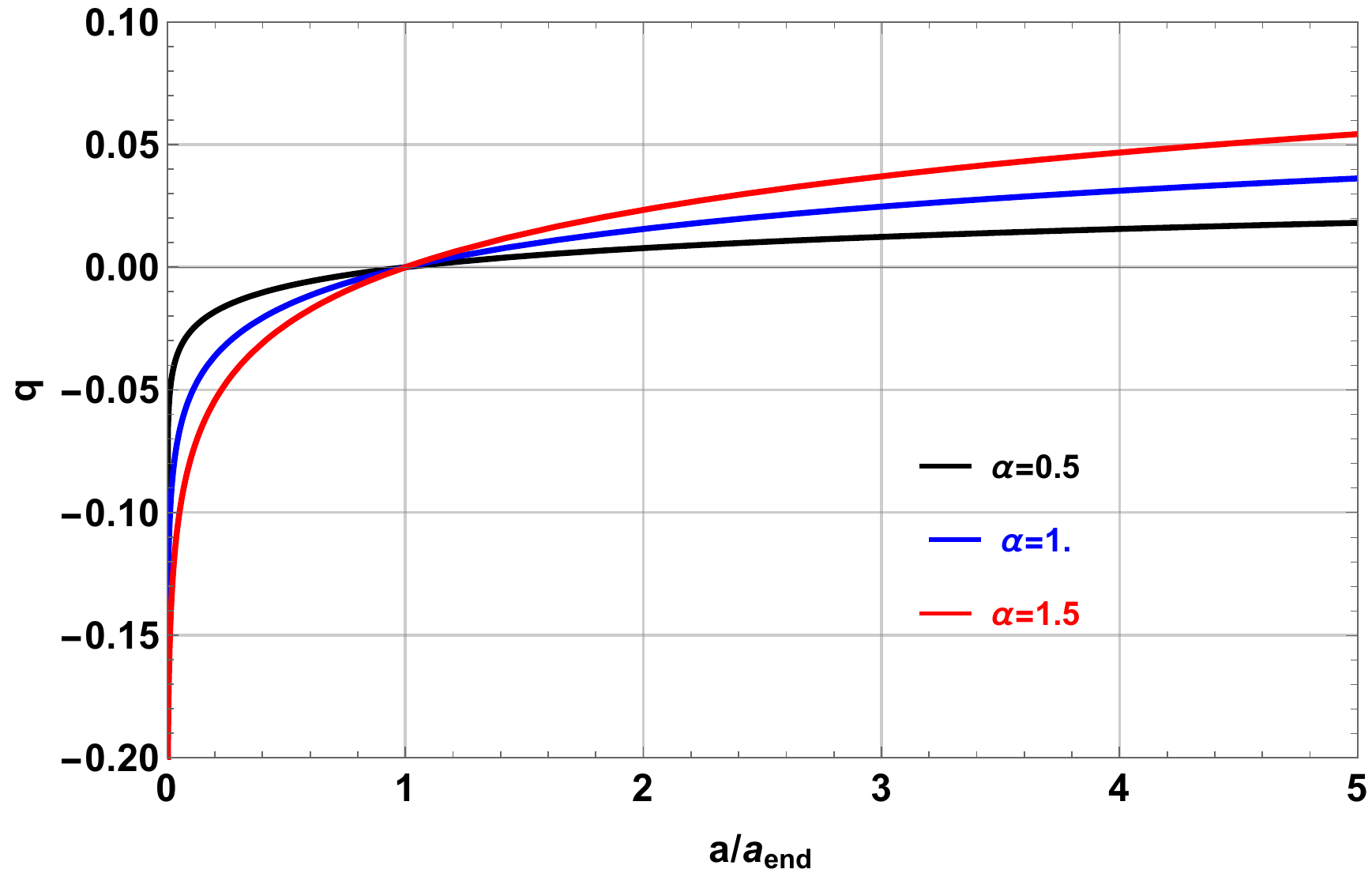} \\
\end{tabular}
\caption{ EoS of state parameter $\omega_B$ and deceleration $q$ as function of the scale factor $a$ for different values of $\alpha$ and $A= -0.985$.}
\label{fig:9}
\end{figure}

In figure \ref{fig:9}, we illustrate the behaviour of the EoS parameter $\omega_B$ and deceleration $q$ as a function of the scale factor $a$ for different values of $\alpha$ and $A= -0.985$.

One aspect of the model is to obtain constraints on the model parameters allowed by current observations, deriving them from the constraints on the scalar spectral index $n_s$ and tensor to scalar ratio $r$. In slow-roll approximation, they can be expressed as \cite{Bamba:2014wda}, 
\begin{equation}
n_s = 1+ 2 \eta - 6\epsilon,  \quad r = 16\epsilon. 
\end{equation}

\begin{figure}
    \centering
    \includegraphics[scale=0.6]{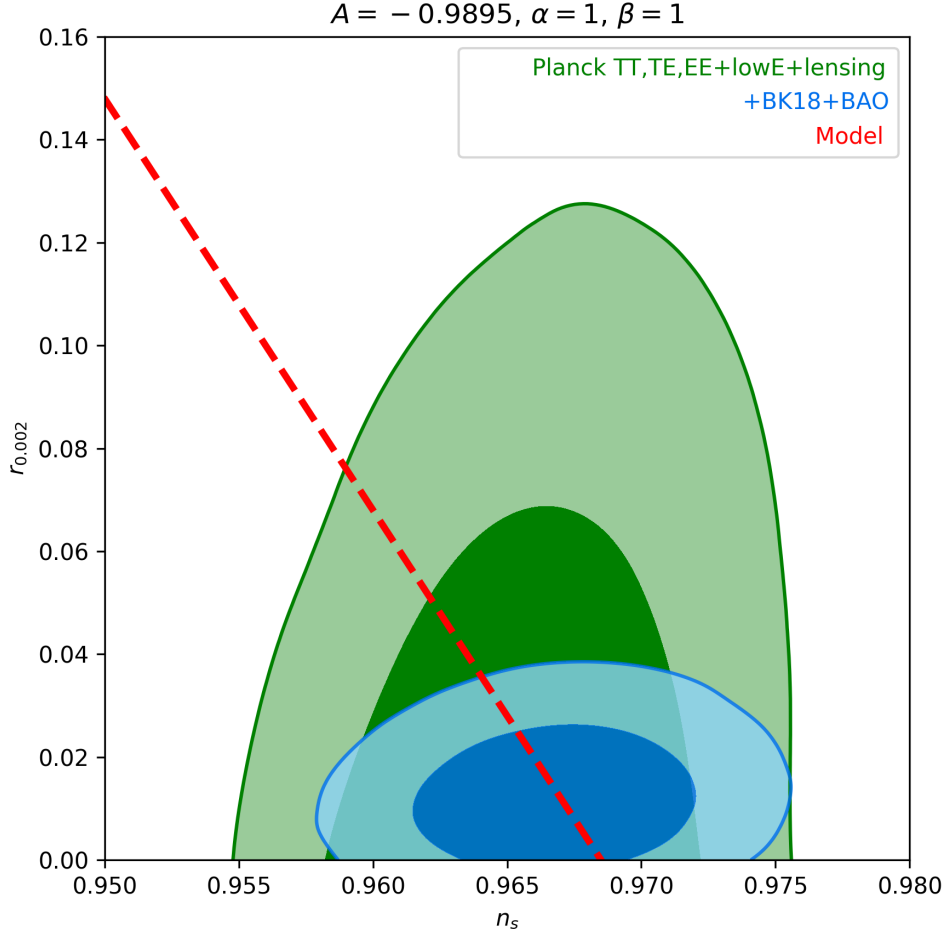}
    \caption{Constraints in the $r$ vs  $n_s$ plane for the Planck 2018 baseline analysis, and when also adding BICEP/Keck data through the end of the 2018 season plus BAO data to
improve the constraint on $n_s$ (taken from \cite{BICEP:2021xfz}). The red dashed line represents a parametric plot of $r$ vs  $n_s$ given by our model \eqref{rvsns} for the parameters $A=-0.9895, \alpha= 1.0, \beta = 1.0$.}
    \label{fig:r-ns}
\end{figure}

In our scenario, the slow-roll parameters defined in  \eqref{eq46}, can be expressed as 
\begin{eqnarray}
& \epsilon(z) =\frac{(z+1) H'(z)}{H(z)}, \quad \eta(z) = \frac{1}{2} \left(\frac{(z+1) H'(z)}{H(z)}+\frac{(z+1) H''(z)}{H'(z)}+1\right), \label{slow-roll-z}
\end{eqnarray}
where, according to the definition of $N$,  \eqref{eqz55}, we have passed to derivatives with respect the redshift through $
    \frac{d f}{d N}=     \frac{d f}{d z}  \frac{d z}{d N} = (1+z) \frac{d f}{d z}$.

For the scalar spectral index $n_s$ and tensor to scalar ratio $r$, we have   
\begin{align}
& n_s(z)= 2 -\frac{5 (z+1) H'(z)}{H(z)}+\frac{(z+1) H''(z)}{H'(z)}, \quad  r(z)= \frac{16 (z+1) H'(z)}{H(z)}.
\end{align}
Replacing $H(z)$ from \eqref{zeq67} in expressions \eqref{slow-roll-z}, we obtain 
\begin{align}
    \epsilon(z) & = 2 + \frac{2 F_{\text{end}} (z+1)^4 f'\left(\frac{F_{\text{end}} (z+1)^4}{( z_{\text{end}}+1)^4}\right)}{( z_{\text{end}}+1)^4
   f\left(\frac{F_{\text{end}} (z+1)^4}{( z_{\text{end}}+1)^4}\right)}, \\
    \eta(z)& = \frac{2 \left(F_{\text{end}} \left(F_{\text{end}} (z+1)^8 f''\left(\frac{F_{\text{end}} (z+1)^4}{(z_{\text{end}}+1)^4}\right)+3
   (z+1)^4 (z_{\text{end}}+1)^4 f'\left(\frac{F_{\text{end}} (z+1)^4}{(z_{\text{end}}+1)^4}\right)\right)+(z_{\text{end}}+1)^8
   f\left(\frac{F_{\text{end}} (z+1)^4}{(z_{\text{end}}+1)^4}\right)\right)}{(z_{\text{end}}+1)^4 \left(F_{\text{end}} (z+1)^4
   f'\left(\frac{F_{\text{end}} (z+1)^4}{(z_{\text{end}}+1)^4}\right)+(z_{\text{end}}+1)^4 f\left(\frac{F_{\text{end}}
   (z+1)^4}{(z_{\text{end}}+1)^4}\right)\right)}.
\end{align}
Using \eqref{eq50}, we acquire 
\begin{align}
  & \epsilon(z)  =   \frac{3 (A+1)}{2 \beta  \left(\frac{F_{\text{end}} (z+1)^4}{(z_{\text{end}}+1)^4}\right)^{\frac{3}{4} \alpha(A+1)}+2}, \quad
   \eta(z) =   \frac{3}{2} (A+1) \left(\frac{\alpha +1}{\beta  \left(\frac{F_{\text{end}}
   (z+1)^4}{(z_{\text{end}}+1)^4}\right)^{\frac{3}{4} \alpha  (A+1)}+1}-\alpha \right), \\
 & n_s(z) = 1 -3 \alpha -3 \alpha  A+\frac{3
   (\alpha -2) (A+1)}{\beta  \left(\frac{F_{\text{end}} (z+1)^4}{(z_{\text{end}}+1)^4}\right)^{\frac{3}{4} \alpha (A+1)}+1}, \quad 
     r(z)  = \frac{24 (A+1)}{\beta  \left(\frac{F_{\text{end}} (z+1)^4}{(z_{\text{end}}+1)^4}\right)^{\frac{3}{4} \alpha (A+1)}+1}, 
\end{align}
which implies 
\begin{equation}
r(n_s)= \frac{8 (3 \alpha  (A+1)+n_s-1)}{\alpha -2}.  \label{rvsns} 
\end{equation}
The Planck 2018 bounds on the spectral index and the tensor-to-scalar ratio  \eqref{eq49}, 
$n_s = 0.9649 \pm 0.0042, r < 0.064$ gives the parameter region $\{-0.992367<A\leq -0.9897, 0.008 (-375 A-251)<n_s<0.9691\}$, or
   $\{-0.9897<A\leq -0.989567, 0.008 (-375 A-251)<n_s<-3 A-2\}$ or
   $\{-0.989567<A<-0.9869, 0.9607<n_s<-3 A-2\}$.

Figure \ref{fig:r-ns} shows constraints in the $r$ vs  $n_s$ plane for the Planck 2018 baseline analysis, and when also adding BICEP/Keck data through the end of the 2018 season plus BAO data to
improve the constraint on $n_s$ (taken from \cite{BICEP:2021xfz}). The red dashed line represents a parametric plot of $r$ vs  $n_s$ given by our model \eqref{rvsns} for the parameters $A=-0.9895, \alpha= 1.0, \beta = 1.0$. The contour displays $1\sigma$ and $2\sigma$ confidence levels from darker to lighter. We have used data from the Planck collaboration + BICEP/Keck + BAO which is public at \url{http://bicepkeck.org}. The red dashed line represents our theoretical model that agrees within $1\sigma$ with the more stringent data.

\section{Phase space analysis}
\label{sec:phase}

In dynamical systems, phase space is a space in which all possible
states of a system are represented. In phase space, every degree of freedom or parameter of the system is represented as an axis of a multidimensional space; a one-dimensional system is called a phase line, while a two-dimensional system is called a phase plane. 
Dynamical systems methods have been proven to be a powerful scheme for investigating the physical behaviour of cosmological models. As we know, there exist four standard ways of systematic investigation that
can be used to examine cosmological models:
   (i)  Obtaining and analyzing exact solutions;
   (ii) Heuristic approximation methods;
   (iii) Numerical simulation, and
   (iv) Qualitative analysis  \cite{Ellis}. 
The last case can be used with three different approaches:
        (a) Piecewise approximation methods,
        (b) Hamiltonian methods,
        (c) Dynamical systems methods.
In approach iv (a), the evolution of the model universe is
approximated through a sequence of epochs in which specific terms
in the governing differential equations can be neglected, leading
to a more straightforward system of equations. This heuristic approach is
firmly based on the existence of heteroclinic sequences, which is
a concept from iv (c). In approach iv (b), Einstein's
equations are reduced to a Hamiltonian system dependent on time
for a particle (point universe) in two dimensions. This approach
has been used mainly for modelling and analyzing the dynamics of
the Universe, nearly the Big Bang singularity (one of the approaches we will follow). In the approach iv
(c) Einstein's equations for homogeneous cosmologies can be
described as an autonomous system of first-order ordinary
differential equations plus certain algebraic constraints. Specifically, Einstein's field equations of Bianchi's
cosmologies and their isotropic subclass (FLRW models) can be
written as an autonomous system of first-order differential
equations whose solution curves partitioned to $\mathbb{R}^ n$ in
orbits, defining a dynamical system in $\mathbb{R}^ n$. In the
general case, singular points, invariant sets, and other  elements of the phase space partition can be listed and
described. This study consists of several steps: determining
singular points, the linearization in a neighbourhood of them, the search for the eigenvalues of the associated Jacobian matrix,
checking the stability conditions in a neighbourhood of the
singular points, the finding of the stability and instability sets
and the determination of the basin of attraction, etcetera. On some
occasions, to do that, it is necessary to simplify a
dynamic system. Two approaches are applied to this objective:
one, reduce the dimensionality of the system, and two, eliminate
the nonlinearity. Two rigorous mathematical techniques that allow
substantial progress along both lines is the centre manifold theory
and normal forms. Using this approach, in \cite{Ellis}, many results have been
obtained  concerning the possible asymptotic
cosmological states in Bianchi and FLRW models, whose material
content is a perfect fluid (usually modelling "dark matter", a component that plays an important role in the formation of
structures in the Universe, such as galaxies and clusters of
galaxies) with linear equation of state (with the possible
inclusion of a cosmological constant). Also, several classes of
inhomogeneous models are examined, comparing the results with those
obtained using numerical and Hamiltonian methods. This analysis is
extended in \cite{Coley:2003mj}, to other contexts, having
considered other material sources, such as the scalar fields.

Moreover, one can use tools of the theory of averaging in nonlinear differential equations and the qualitative analysis of dynamical systems to obtain relevant information about the solution's space of cosmological models. The averaging methods were applied extensively in \cite{Alho:2015cza, Alho:2019pku, Fajman:2020yjb, Fajman:2021cli, Leon:2021lct, Leon:2021rcx} to single field scalar field cosmologies, and for scalar field cosmologies with two scalar fields which interact only gravitationally with the matter in \cite{Chakraborty:2021vcr}.
Within this context, one deal with perturbation problems of differential systems expressed in Fenichel's normal form \cite{dumortier, Fenichel, Fusco, Berglund, holmes, Kevorkian1, Verhulst}. That is, given $(x,y)\in \mathbb{R}^{n+m}$ and $f, g$ smooth functions, equations can be written as: 
\begin{equation}
\label{eq:1.1}
\dot x=f(x,y; \varepsilon), \quad \dot y=\varepsilon g(x,y; \varepsilon), \quad x=x(t), \quad y=y(t).
\end{equation}
The system \eqref{eq:1.1} is called ``fast system''  as opposed to 
\begin{equation}
\label{eq:1.2}
\varepsilon  x'=f(x,y; \varepsilon), \quad y'=g(x,y; \varepsilon), \quad x=x(\tau), \quad y=y(\tau),
\end{equation}
which is obtained after the re scaling $\tau=\varepsilon t$, and  is called ``slow system''.  

Notice that for $\varepsilon>0$, the phase portraits of \eqref{eq:1.1} and \eqref{eq:1.2} coincide.  It follows two problems that manifestly depend on two scales: (i) the problem in terms of the "slow time" variable, whose solution is analogous to the outer solution in a boundary layer problem; (ii) the fast system: a change of scale on the system which describes the rapid evolution that occurs in shorter times; analogous to the inner solution of a boundary layer problem.
The solution of each subsystem will be sought in the form of a regular perturbation expansion. The subsystems will have
simpler structures for singularly perturbed problems than the complete problems. Then, the slow and fast dynamics are characterized by reduced phase line or phase plane dynamics. Therefore, information on the dynamics for small values of $\varepsilon$ is obtained. This technique is used to construct uniformly valid approximations of the solutions of perturbation problems using seed solutions that satisfy the original equations in the limit of $\varepsilon\rightarrow 0$ \cite{holmes}.

\subsection{Phase space analysis: pure NLED}
\label{Sec. 4}
The equation (\ref{eq64}) represents the motion of a particle of the unit mass in the effective potential. 
This equation is satisfied on the zero-energy level, where $\rho_{B}$ plays the role of effective energy density parameterized through the scale factor $a(t)$. Therefore the standard cosmological model can be represented in terms of a dynamical system of a Newtonian type: 
\begin{equation}
\ddot a=-\frac{\partial V_{\text{eff}}}{\partial a}, \quad V_{\text{eff}} \left( a \right)=- \frac{{F}_{\text{end}} ~a_{\text{end}}^4}{6 a^2} ~f \left({F}_{\text{end}} \left(\frac{a_{\text{end}}}{a} \right)^4\right). 
\label{eq70}
\end{equation} 
The scale factor $a$ plays the role of a positional variable of a fictitious unit mass particle, miming the Universe's expansion.  

We assume ${F}_{\text{end}} > 0$, and introduce the variables 
\begin{equation}
\frac{a}{a_{\text{end}}} =e^u = e^{-N} , \quad v=\frac{\dot a}{a_{\text{end}}} ~\frac{1}{\sqrt{2 {{F}_{\text{end}}}}},
\label{eq71}
\end{equation}
 and the time variable
\begin{equation}
\tau=  \sqrt{2 {F}_{\text{end}}} ~\int e^{-u} dt. 
\label{eq72}
\end{equation}
This system can be written in the form 
\begin{equation}
\frac{d u}{d\tau}= v, \quad  \frac{d v}{d \tau}= -\frac{\partial W(u)}{\partial u}, 
\label{eq73}
\end{equation}
with effective particle-potential 
\begin{equation}
W(u)=-\frac{1}{12} e^{-2 u} ~f \left({F}_{\text{end}} ~e^{-4 u}\right).
\label{eq74}
\end{equation}
Thus,  $\frac{v^2}{2}+W(u)=E$, is the constant of energy. 
From the above system, we see that, generically, the equilibrium points of the system \eqref{eq73} are situated on the axis $u$ ($v=0$), and they satisfy $\frac{\partial W(u)}{\partial u}=0$. From the characteristic equation, it follows that just three types of equilibrium points are admitted: 

\begin{enumerate}
\item Saddle if $u_c: \frac{\partial W}{\partial u}|_{u=u_c}=0$ and $\frac{\partial^2 W}{\partial u^2}|_{u=u_c}<0;$ 
\item Focus if $u_c: \frac{\partial W}{\partial u}|_{u=u_c}=0$ and $\frac{\partial^2 W}{\partial u^2}|_{u=u_c}>0;$ 
\item Degenerated critical point if  $u_c: \frac{\partial W}{\partial u}|_{u=u_c}=0$ and $\frac{\partial^2 W}{\partial u^2}|_{u=u_c}=0.$ 
\end{enumerate}
We have the expressions 

\begin{eqnarray}
W'(u) & = & \frac{1}{3} {F}_{\text{end}}  ~e^{-6 u} f'\left({F}_{\text{end}} ~e^{-4 u}\right) + 
\frac{1}{6} e^{-2 u} f\left({F}_{\text{end}} ~e^{-4 u}\right), \nonumber \\
W''(u) & = & -\frac{4}{3} {F}_{\text{end}}^2 ~e^{-10 u} f'' \left({F}_{\text{end}} ~e^{-4 u}\right) -
\frac{8}{3} {F}_{\text{end}} ~e^{-6 u} f' \left({F}_{\text{end}} ~e^{-4 u}\right) - 
\frac{1}{3} e^{-2 u} f \left({F}_{\text{end}}  ~e^{-4 u}\right).
\label{eq75}
\end{eqnarray}

The eigenvalues of the Jacobian matrix evaluated at the equilibrium point with coordinate $u_c$ are \newline $\left\{-i \sqrt{W'' (u_c)}, i \sqrt{W'' (u_c)}\right\}$, such that the condition for having periodic solutions is $W'' (u_c)>0$. Since $W' (u_c)=0$ at the equilibrium points, we end up with the condition  
$2 {F}_{\text{end}} ~f''\left({F}_{\text{end}} ~e^{-4 u_c}\right)+3 e^{4 u_c} f' \left({F}_{\text{end}} ~e^{-4 u_c}\right)<0$ as a sufficient condition for having a cyclic Universe. 

Due to the relation 
${F}= {F}_{\text{end}} ~e^{-4 u}$, the above conditions can be summarized as follows:

\begin{enumerate}
\item The equilibrium points are given by $u_c=1/4\ln({F}_{\text{end}}/{F}_c)$, where

\begin{equation}
2 {F}_c f' \left({F}_c \right)+  f \left({F}_c \right) =0. 
\label{eq76}
\end{equation}
The above equation must be considered an algebraic (in most cases transcendent) equation of ${F}_c$ for a given $f$ and not a differential equation for $f$. 

Furthermore, evaluated at the equilibrium point, we obtain $\rho_{B}+3 p_{B}=0$. That is zero acceleration point. That is not unexpected since the condition for obtaining the equilibrium points is $\ddot a=0$. Evaluating the equilibrium point, we obtain 
$c_s^2 |_{u=u_c} =-\frac{4}{3} \frac{f'' \left({F}_c \right)}{f' \left({F}_c \right)}-\frac{7}{3}$.
The conditions for classical stability at the equilibrium points will be 
\begin{equation}
\frac{7}{4}\leq -\frac{f'' \left({F}_c \right)}{f' \left({F}_c \right)}<\frac{5}{2}.
\label{eq77}
\end{equation}
\item The equilibrium point is a saddle for 
\begin{equation}
4 {F}_c^2 f'' \left({F}_c \right) + 8 {F}_c f' \left({F}_c \right) + f \left({F}_c \right) > 0,
\label{eq78}
\end{equation}
or, equivalently:
\begin{eqnarray}
2 {F}_{\text{end}}^2  f'' \left({F}_c \right)+ 3 {F}_c f' \left({F}_c \right) > 0.
\end{eqnarray}
 
\item The equilibrium point is a focus for 
\begin{equation*}
\frac{\partial^2 W}{\partial u^2}|_{u=u_c} > 0, 
\end{equation*}
or, equivalently:
\begin{eqnarray}
2 {F}_{\text{end}}^2  f'' \left({F}_c \right)+ 3 {F}_c f' \left({F}_c \right) < 0. 
\label{eq79}
\end{eqnarray}
This condition leads to the existence of periodic solutions.  
\item It is degenerate for 
\begin{equation}
f'({F}_c)= -\frac{f({F}_c)}{2 {F}_c}, \quad f''({F}_c)=\frac{3 f({F}_c)}{4 {F}_c^2}.
\label{eq80}
\end{equation}
\end{enumerate}

Using (\ref{eq50}), the effective potential is written as  
\begin{eqnarray}
    W(u) & = &  -\frac{1}{12} {F}_{\text{end}}^{\frac{3 A}{4}-\frac{1}{4}} e^{-(3 A+1) u} \left(\beta  {F}_{\text{end}}^{\frac{3}{4} \alpha  (A+1)} e^{-3 \alpha  (A+1) u}+1\right)^{-1/\alpha }. 
		\label{eq82}
\end{eqnarray}
System (\ref{eq73}) becomes 
\begin{align}
 & \frac{d u}{d\tau}=  v, \\
 & \frac{d v}{d\tau}= \frac{1}{12} {F}_{\text{end}}^{\frac{1}{4} (3 A-1)} e^{-u (3 \alpha +3 (\alpha +1) A+1)} \left(2 \beta  {F}_{\text{end}}^{\frac{3}{4} \alpha  (A+1)}-(3 A+1) e^{3 \alpha  (A+1) u}\right) \left(\beta 
   {F}_{\text{end}}^{\frac{3}{4} \alpha  (A+1)} e^{-3 \alpha  (A+1) u}+1\right)^{-\frac{1}{\alpha }-1}.
\end{align}

The equilibrium points of the system (\ref{eq73})
are $(u_c, 0)$ such that $u_c=1/4\ln({F}_{\text{end}}/{F}_c)$ where ${F}_c$ are the roots  of (\ref{eq76}) which is reduced to 
\begin{equation}
{F}_{c}^{\frac{1}{4} (3 A+1)} \left(2 \beta  {F}_{c}^{\frac{3}{4} \alpha  (A+1)}-3 A-1\right)\left(\beta  {F}_{c}^{\frac{3}{4} \alpha  (A+1)}+1\right)^{-\frac{\alpha +1}{\alpha }} =0. \end{equation}
We assume ${F}_c\neq 0$, then, we have either 
$ {F}_c^{\frac{3}{4} \alpha 
   (A+1)}= \beta^{-1}(1+3 A)/2$ (for all values of $\alpha$ and $(1+3 A)/\beta>0$) or $ {F}_c^{\frac{3}{4} \alpha 
   (A+1)} = -\beta^{-1}$ (provided $-1< \alpha <0, \beta<0$). That is, ${F}_c= \left[\beta^{-1}(1+3 A)/2\right]^{\frac{4}{3\alpha(A+1)}}$ (for all values of $\alpha$ and $(1+3 A)/\beta>0$)  or ${F}_c = (-\beta)^{-\frac{4}{3\alpha(A+1)}}$ (provided $-1< \alpha <0, \beta<0$). 
Hence, the equilibrium points are 
\begin{equation}
P_0:=\Bigg(\frac{\ln (2)}{3 \alpha (1+ A)}+\frac{1}{4} \ln \left[{F}_{\text{end}} \left(\frac{3 A+1}{\beta }\right)^{-\frac{4}{3 \alpha(1+A)}}\right], 0\Bigg),  \quad P_1:=\Bigg(\frac{1}{4} \Bigg[\ln ({F}_{\text{end}})-\frac{4 \ln
   \Big(-\frac{1}{\beta }\Big)}{3 \alpha  (A+1)}\Bigg], 0\Bigg). 
\end{equation}
$P_0$ exists for $(1+3 A)/\beta>0$ and 
$P_1$ exists for $-1 <\alpha < 0, \beta <0$. From physical conditions, we assume $\beta\geq 0$. That is,  the equilibrium point $P_1$ is discarded. Then, for $P_0$ does exists we require $A>-1/3,  \beta>0$. 

The linearization matrix evaluated at $P_0$ has eigenvalues
\begin{align*}
&  -\frac{i 3^{-\frac{\alpha +1}{2 \alpha }} \sqrt{\alpha } 2^{\frac{1}{3 \alpha +3 \alpha  A}-\frac{1}{2}} (A+1)^{-\frac{1}{2 \alpha} } (3 A+1)^{\frac{1}{6} \left(\frac{3
   A+1}{\alpha +\alpha  A}+3\right)} \beta ^{-\frac{3 A+1}{6 \alpha +6 \alpha  A}}}{\sqrt[4]{{F}_{\text{end}}}}, \\
   & \frac{i 3^{-\frac{\alpha +1}{2 \alpha }} \sqrt{\alpha } 2^{\frac{1}{3 \alpha +3
   \alpha  A}-\frac{1}{2}} (A+1)^{-\frac{1}{2 \alpha} } (3 A+1)^{\frac{1}{6} \left(\frac{3 A+1}{\alpha +\alpha  A}+3\right)} \beta ^{-\frac{3 A+1}{6 \alpha +6 \alpha 
   A}}}{\sqrt[4]{{F}_{\text{end}}}}. 
\end{align*}
For $\alpha<0$, $A>-1/3,  \beta>0$, the eigenvalues are reals of different signs, then $P_0$ is a saddle. If  $\alpha>0$,  $A>-1/3,  \beta>0$, the eigenvalues are purely imaginary, and the point is nonhyperbolic.

On the other hand, using the generating function $f$ given by  is  \eqref{eq50},  condition for   classical stability and causality is \eqref{eq77} for $P_0$ is 
\begin{equation}
\frac{7}{4}\leq -2^{\frac{4}{3 \alpha +3 \alpha  A}-2} (3 A+1)^{-\frac{4}{3 \alpha +3 \alpha  A}} (\alpha +3 \alpha  A-6) \beta ^{\frac{4}{3 \alpha +3 \alpha  A}}<\frac{5}{2}. \label{EQ89}
\end{equation}
Then, assuming  $\alpha>0$,  $A>-1/3, \beta>0$, we have that $P_0$ can be a nonlinear centre or nonlinear spiral (because it is nonhyperbolic). The classical stability and causality condition is also given by \eqref{EQ89}. Due to the non-hyperbolicity, we use numerical methods to investigate the stability.

For the numerics it is convenient to use the variables $({F}, v)$ through the redefinition 
 $u =1/4\ln({F}_{\text{end}}/{F} )$. 
 Hence, 
 \begin{align}
     & \frac{d{F}}{d\tau}= -4 v {F}, \label{eq89}\\
     &  \frac{d{v}}{d\tau} =\varepsilon \left\{{F}^{\frac{1}{4} (3 A+1)} \left(\beta  {F}^{\frac{3}{4} \alpha  (A+1)}+1\right)^{-\frac{\alpha +1}{\alpha }}
   \left(2 \beta  {F}^{\frac{3}{4} \alpha  (A+1)}-3 A-1\right)\right\},  \label{eq90}
 \end{align}
 where 
 \begin{equation}
     \varepsilon:= \frac{1}{12 \sqrt{{F}_{\text{end}}}}.
 \end{equation}
According to our previous estimations,  for the theoretical prior $z_{\text{end}}\simeq 10^{-2} z_{GUT}$,   we have $5 \times 10^{23} \mathrm{cm}^{-2}\lesssim {F}_{\text{end}} \lesssim 5 \times 10^{37} \mathrm{cm}^{-2} $.  Then,  we have
$1.17851\times 10^{-20} \mathrm{cm} \lesssim \varepsilon:= \frac{1}{12 \sqrt{{F}_{\text{end}}}}  \lesssim 1.17851\times 10^{-13} \mathrm{cm}$, therefore, we are in the presence of a fast-slow system. 
The system  \eqref{eq89} and \eqref{eq90} gives the dynamics in the fast manifold. That corresponds to the ``horizontal motion'' $v=v_0$ (constant), and 
\begin{equation}
  \frac{d{F}}{d\tau}= -4 v_0 {F} \implies {F}(\tau)= {F}(\tau_0) e^{-4 v_0 (\tau -\tau_0)} .  
\end{equation}
This fact is confirmed numerically by  considering the parameters
\begin{align}
& \alpha=1,  \quad \beta=1,  \quad A= -0.9895, \quad {F}_{\text{end}} \in\{5 \times 10^{23}, 
5 \times 10^{37}\}. \label{param1}
\end{align}

\begin{figure}[ht!]
    \centering
    \includegraphics[scale=0.6]{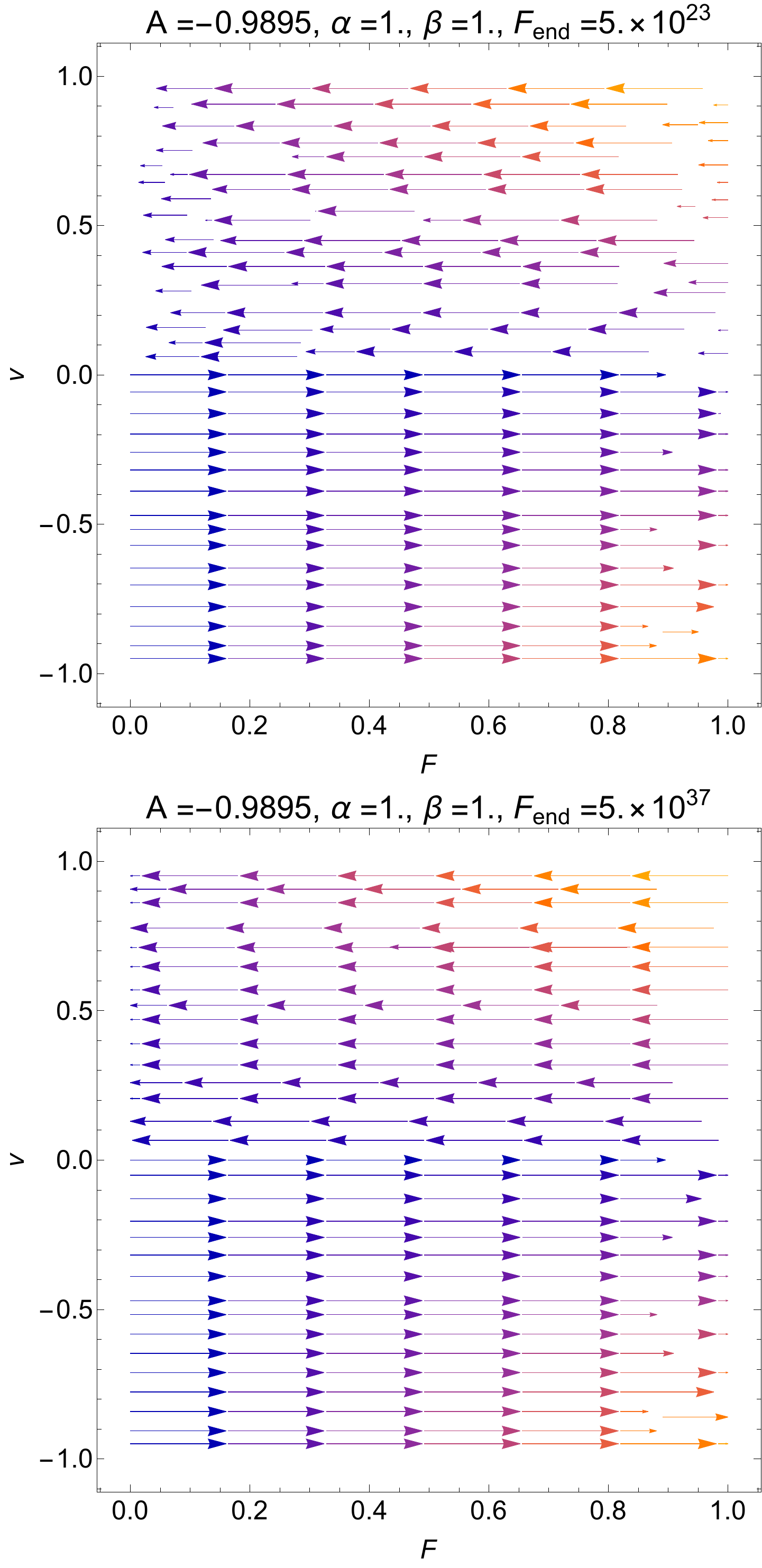}
    \caption{Horizontal flow of the system  \eqref{eq89} and \eqref{eq90} for some values in the parameter space \eqref{param1}.}
    \label{fig:my_label}
\end{figure}

Using the slow time $T= \varepsilon\tau$, we have the system 
 \begin{align}
     & \varepsilon \frac{d{F}}{dT}= -4   v {F}, \label{Redeq89}\\
     & \frac{d{v}}{dT} = {{F}^{\frac{1}{4} (3 A+1)} \left(\beta  {F}^{\frac{3}{4} \alpha  (A+1)}+1\right)^{-\frac{\alpha +1}{\alpha }}
   \left(2 \beta  {F}^{\frac{3}{4} \alpha  (A+1)}-3 A-1\right)}.  \label{Redeq90}
 \end{align}
We can easily see that the equilibrium governs the dynamics at the slow manifold points $({F}, v)$, which satisfies $v=0$ and $\left(\beta  {F}^{\frac{3}{4} \alpha  (A+1)}+1\right)^{-\frac{\alpha +1}{\alpha }}
   \left(2 \beta  {F}^{\frac{3}{4} \alpha  (A+1)}-3 A-1\right)=0$. Depending on the parameter values, the attractors are $P_0$ or $P_1$. They were analyzed in the coordinates $(u,v)$. Notice that that the axis $({F},v)=(0,v_c)$ is a line of equilibrium points, with eigenvalues $\{0,-4 v_c\}$.

\subsection{Phase space analysis: NLED including matter}
\label{Sec. 5}

For the metric  \eqref{eq13}, the Friedmann equations for NLED with an extra matter source (a matter fluid with density $\rho_m$, pressure $p_m$ and a barotropic equation of state $p_m=w_m \rho_m$) can be easily computed which results in, 
\begin{eqnarray}
H^2 & = & \left( \frac{\dot{a}}{a} \right)^2 = \frac{1}{3} ~ (\rho_B + \rho_m), \nonumber \\
3 \frac{\ddot a}{a} & = & - \frac{1}{2} \left( \rho_B + 3 p_B + \rho_m + 3 p_m\right),
\label{matterB}
\end{eqnarray}
where $H = \dot{a}/a$ is the Hubble parameter.

The conservation of the energy-momentum tensor $\nabla ^{\mu }T_{\mu \nu }=0$ leads to the continuity equation of the NLED being given by \eqref{eq15}.
The continuity equation for barotropic matter is given by
\begin{equation}
  \dot{\rho_m}+ 3 (1+w_m)H\rho_m=0.  \label{matter_rho}
\end{equation}
Therefore, 
\begin{eqnarray}
\rho_m & = &\rho_{m, end} \left(\frac{a_{\text{end}}}{a} \right)^{3(1+w_m)}= \rho_{m, end} \left(\frac{1+z}{1+ z_{\text{end}}}\right)^{3(1+w_m)},
\label{eq91}
\end{eqnarray}
where 
\begin{equation}
    \rho_{m, end}= 3 H_{\text{end}}^2 -\rho_{\text{end}}, \quad \rho_{\text{end}}= {F}_{\text{end}} f({F}_{\text{end}}).
\end{equation}
We have \begin{eqnarray}
 q(z) & = & -1+\frac{(1+z)}{2} \frac{d \ln H^2(z)}{d z},
\end{eqnarray}
where 
\begin{eqnarray}
    H^2(z) & = & \frac{1}{3} ~ \left\{{F}_{\text{end}}\; \left(\frac{1+z}{1+ z_{\text{end}}}\right)^4 \; f \left({F}_{\text{end}}\left[\frac{1+z}{1+ z_{\text{end}}}\right]^4\right)   +  \rho_{m, end} \left(\frac{1+z}{1+ z_{\text{end}}}\right)^{3(1+w_m)}\right\}. \label{zeq111}
\end{eqnarray}

We assume ${F}_{\text{end}} > 0$, and using the variables \eqref{eq71},  and the time variable $\tau$ given by \eqref{eq72}, 
the system is then equivalent to 
\begin{align}
&\frac{d u}{d\tau}= v, \quad \frac{d v}{d \tau}= -\frac{\partial W(u)}{\partial u},
\end{align}
where the effective potential is
\begin{equation}
W(u)=-\frac{1}{12}  \frac{\rho_{m, end}}{{F}_{\text{end}}} e^{-u (3 w_m+1)}
-\frac{1}{12} e^{-2 u}   f\left({F}_{\text{end}} e^{-4 u}\right).\label{eq93}
\end{equation}
Now, the equilibrium points are found by solving numerically 
\begin{align}
\label{Px}
W'(u)=0. 
\end{align}

As in the previous section, a given equilibrium point $u_c$ is of one of the following types: 
\begin{enumerate}
\item saddle if $u_c: \frac{\partial W}{\partial u}|_{u=u_c}=0$ and $\frac{\partial^2 W}{\partial u^2}|_{u=u_c}<0;$ 
\item focus if $u_c: \frac{\partial W}{\partial u}|_{u=u_c}=0$ and $\frac{\partial^2 W}{\partial u^2}|_{u=u_c}>0;$ 
\item degenerated critical point if  $u_c: \frac{\partial W}{\partial u}|_{u=u_c}=0$ and $\frac{\partial^2 W}{\partial u^2}|_{u=u_c}=0.$ 
\end{enumerate}
We have the expressions 
\begin{align}
&W'(u)=\frac{1}{3} {F}_{\text{end}} e^{-6 u} f'\left({F}_{\text{end}}
   e^{-4 u}\right)+\frac{1}{6} e^{-2 u}
   f\left({F}_{\text{end}} e^{-4
   u}\right)+\frac{\rho_{m, end} (3 w_m+1)
   e^{-u (3 w_m+1)}}{12 {F}_{\text{end}}},\\
& W''(u)=-\frac{4}{3} {F}_{\text{end}} e^{-10 u}
   \left({F}_{\text{end}} f''\left({F}_{\text{end}} e^{-4
   u}\right)+2 e^{4 u} f'\left({F}_{\text{end}} e^{-4
   u}\right)\right)  -\frac{1}{3} e^{-2 u}
   f\left({F}_{\text{end}} e^{-4
   u}\right) \nonumber \\
   & -\frac{\rho_{m, end} (3 w_m+1)^2
   e^{-u (3 w_m+1)}}{12 {F}_{\text{end}}}.
\end{align}

As before, the eigenvalues of the Jacobian matrix evaluated at the equilibrium point with coordinate $u_c$ are \\$\left\{-i \sqrt{W'' (u_c)}, i \sqrt{W'' (u_c)}\right\}$, such that the condition for having periodic solutions is $W'' (u_c)>0$. Since $W'(u_c)=0$ at the equilibrium points, $u_c$ satisfies the equation \; 
$\frac{2 {F}_{\text{end}} e^{u_c (3 w_m-5)}
   \left(2 {F}_{\text{end}} f'\left({F}_{\text{end}} e^{-4
   u_c}\right)+e^{4 u_c}
   f\left({F}_{\text{end}} e^{-4
   u_c}\right)\right)}{3
   w_m+1}+ \rho_{m, end}=0$, \\
   we end up with the condition 
\\
$2 {F}_{\text{end}} e^{2 u_c}
   \left(e^{4 u_c} (3
   w_m-7)
   f'\left({F}_{\text{end}} e^{-4
   u_c}\right)-4 {F}_{\text{end}}
   f''\left({F}_{\text{end}} e^{-4
   u_c}\right)\right)+e^{10
   u_c} (3 w_m-1)
   f\left({F}_{\text{end}} e^{-4
   u_c}\right)>0$ \\as a sufficient condition for having a cyclic universe.

Due to the relation 
${F}={F}_{\text{end}} e^{-4 u}$, the above conditions can be summarized as follows:

\begin{enumerate}
\item The equilibrium points are given by $u_c=1/4\ln({F}_{\text{end}}/{F}_{c})$, where ${F}_{c}$ are the zeroes of the algebraic (most of the cases transcendent) equation 
\begin{equation}
 {2 {F}_c \left(2
   {F}_c
   f'({F}_c)+f({F}_c)\right)
   +(3
   w_m+1) \rho_{m, end} \left(\frac{{F}_c}{{F}_{\text{end}}}\right)^{\frac{3
   (w_m+1)}{4}}}=0,
\end{equation}
for a given $f$. 

\item The equilibrium point is a saddle for 
\begin{eqnarray}
&&\left(2 {F}_{c} \left(4
   {F}_{c} f''({F}_{c})+(7-3
   w_m)
   f'({F}_{c})\right)+(1-3
   w_m)
   f({F}_{c})\right)>0. 
\end{eqnarray}

\item The equilibrium point is a focus for 
\begin{eqnarray}
&&\left(2 {F}_{c} \left(4
   {F}_{c} f''({F}_{c})+(7-3
   w_m)
   f'({F}_{c})\right)+(1-3
   w_m)
   f({F}_{c})\right)<0.
\end{eqnarray}
This condition leads to the existence of periodic solutions.  
\item It is degenerated for
\begin{align}
&f'({F}_{c})= 
   -\frac{2 {F}_{c}
   f({F}_{c})+\rho_{m, end}(1+3  w_m)
   \left(\frac{{F}_{\text{end}}}{{F}_{c}}\right)^{-\frac{3}{4}
   (w_m+1)}}{4
   {F}_{c}^2}, \\
   & f''({F}_{c})=
   \frac{12 {F}_{c}
   f({F}_{c})+ \rho_{\text{end}}
   (7-9 (w_m-2) w_m)
   \left(\frac{{F}_{\text{end}}}{{F}_{c}}\right)^{-\frac{3}{4}
   (w_m+1)}}{16
   {F}_{c}^3}.
\end{align}
\end{enumerate}

We apply the procedure to the present model \eqref{eq50}. That is, using (\ref{eq50}), the effective potential is written as  
\begin{eqnarray}
    W(u) & = & -\frac{1}{12}  \frac{\rho_{m, end}}{{F}_{\text{end}}} e^{-u (3 w_m+1)}  -\frac{1}{12} {F}_{\text{end}}^{\frac{3 A}{4}-\frac{1}{4}} e^{-(3 A+1) u} \left(\beta  {F}_{\text{end}}^{\frac{3}{4} \alpha  (A+1)} e^{-3 \alpha  (A+1) u}+1\right)^{-1/\alpha }. 
		\label{Beq82}
\end{eqnarray}
System (\ref{eq73}) becomes 
\begin{align}
 & \frac{d u}{d\tau}=  v, \\
 & \frac{d v}{d\tau}= -\frac{\rho_{m, end} (3 w_m+1)
   e^{-u (3 w_m+1)}}{12 {F}_{\text{end}}} \nonumber \\
 & +  \frac{1}{12} {F}_{\text{end}}^{\frac{1}{4} (3 A-1)} e^{-u (3 \alpha +3 (\alpha +1) A+1)} \left(2 \beta  {F}_{\text{end}}^{\frac{3}{4} \alpha  (A+1)}-(3 A+1) e^{3 \alpha  (A+1) u}\right) \left(\beta 
   {F}_{\text{end}}^{\frac{3}{4} \alpha  (A+1)} e^{-3 \alpha  (A+1) u}+1\right)^{-\frac{1}{\alpha }-1}.
\end{align}
   For the numerics it is convenient to use the variables $({F}, v)$ through the redefinition 
 $u =1/4\ln({F}_{\text{end}}/{F} )$. 
 Hence, 
 \begin{align}
     & \frac{d{F}}{d\tau}= -4 v {F}, \label{Neq89}\\
     &  \frac{d{v}}{d\tau} = \varepsilon \left\{-  \frac{\rho_{m, end}}{ {{F}_{\text{end}}}^{\frac{3 (w_m+1)}{4}}  } (3 w_m+1)    {F} ^{\frac{3 w_m +1}{4}}  + {{F}^{\frac{1}{4} (3 A+1)} \left(\beta  {F}^{\frac{3}{4} \alpha  (A+1)}+1\right)^{-\frac{\alpha +1}{\alpha }}
   \left(2 \beta  {F}^{\frac{3}{4} \alpha  (A+1)}-3 A-1\right)} \right\}, \label{Neq90}
 \end{align}
  where 
 \begin{equation}
     \varepsilon:= \frac{1}{12 \sqrt{{F}_{\text{end}}}}.
 \end{equation}
According to our previous estimations,  for the theoretical prior $z_{\text{end}}\simeq 10^{-2} z_{GUT}$,   we have $5 \times 10^{23} \mathrm{cm}^{-2}\lesssim {F}_{\text{end}} \lesssim 5 \times 10^{37} \mathrm{cm}^{-2} $. Furthermore,  from equations \eqref{eq91} and \eqref{eq43}, we have 
\begin{eqnarray}
\rho_{m, end} & = & \rho_{m,0} \left(\frac{a_0}{a_{\text{end}}} \right)^{3(1+w_m)}= 3 H_0^2 \Omega_{m,0} \left(1+z_{\text{end}}\right)^{3(1+w_m)}.
\end{eqnarray}
We consider dust matter ($w_m=0$) and assume $\beta=1$. Next, using $H_{0}=h\,1.08\times 10^{-30}\mathrm{cm}^{-1}$, where  $h =(67.4\pm 0.5)\times 10^{-2}$,  $\Omega_{m0}=0.315\pm 0.007$ and $N_{\text{eff}}=2.99 \pm 0.17$ according to the Planck 2018 results \cite{Planck2018}, and considering the theoretical prior $z_{\text{end}}\simeq 10^{-2} z_{GUT}$, we have 
$4.8\times 10^{17} \mathrm{cm}^{-2} \lesssim \rho_{m, end} \lesssim 5.2 \times 10^{17} \mathrm{cm}^{-2}$.
We select the best-fit value
$\rho_{m, end} \simeq 5.0\times 10^{17}$.  Then,  we have
$1.17851\times 10^{-20} \mathrm{cm} \lesssim \varepsilon:= \frac{1}{12 \sqrt{{F}_{\text{end}}}}  \lesssim 1.17851\times 10^{-13} \mathrm{cm}$, and for $w_m=0$,  we have  $2.65915 \times 10^{-11} \lesssim \frac{\rho_{m, end}}{ {{F}_{\text{end}}}^{\frac{3 (w_m+1)}{4}}  } \lesssim 0.840896$ therefore, we are in the presence of a fast-slow system.  
As before, \eqref{Neq89} and \eqref{Neq90}
gives the dynamics in the fast manifold. That corresponds to the ``horizontal motion'' $v=v_0$ (constant), and $ {F}(\tau)= {F}(\tau_0) e^{-4 v_0 (\tau -\tau_0)}$. 
This fact is confirmed numerically by  considering the parameters 
\begin{align}
& \alpha=1,  \quad \beta=1,  \quad A= -0.9895,  \quad w_m=0,   \quad \rho_{m, end} = 5.0\times 10^{17}, \quad {F}_{\text{end}} \in\{5 \times 10^{23}, 5 \times 10^{37}\}. \label{param2}
\end{align}
\begin{figure}[ht!]
    \centering
    \includegraphics[scale=0.45]{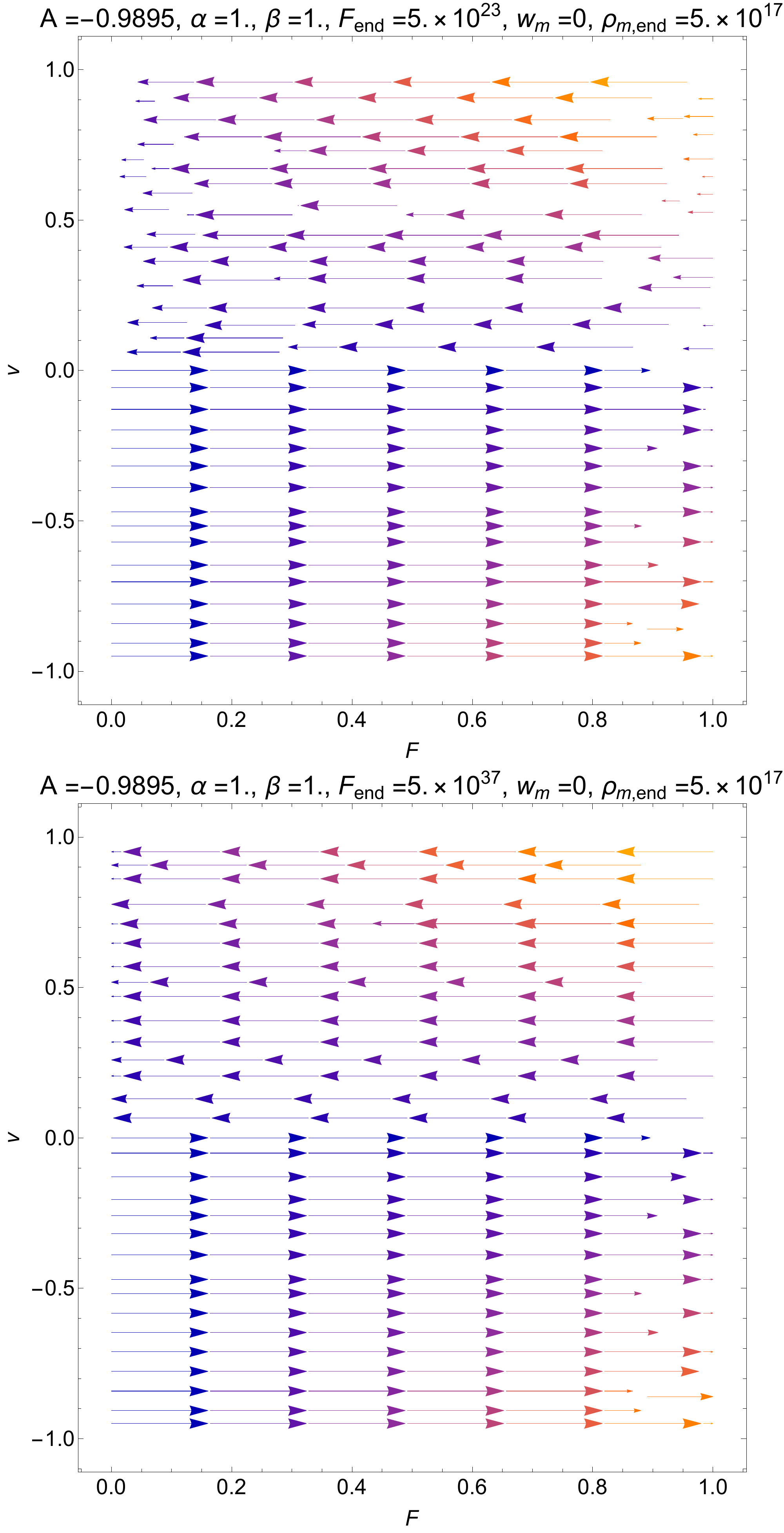}
    \caption{Horizontal flow of the  \eqref{Neq89} and \eqref{Neq90} for some values in the parameter space \eqref{param2}.}
    \label{fig:my_labelDS2}
\end{figure}
 In figure \ref{fig:my_labelDS2}  is presented the horizontal flow of the system \eqref{Neq89} and \eqref{Neq90}  for some values in the parameter space \eqref{param2}.

Using the slow time $T= \varepsilon\tau$, we have the system 
 \begin{align}
     & \varepsilon \frac{d{F}}{dT}= -4   v {F}, \label{Redeq126}\\
     & \frac{d{v}}{dT} = -  \frac{\rho_{m, end}}{ {{F}_{\text{end}}}^{\frac{3 (w_m+1)}{4}}  } (3 w_m+1)    {F} ^{\frac{3 w_m +1}{4}}  +  {{F}^{\frac{1}{4} (3 A+1)} \left(\beta  {F}^{\frac{3}{4} \alpha  (A+1)}+1\right)^{-\frac{\alpha +1}{\alpha }}
   \left(2 \beta  {F}^{\frac{3}{4} \alpha  (A+1)}-3 A-1\right)}.  \label{Redeq127}
 \end{align}
We can easily see that the equilibrium governs the dynamics at the slow manifold points $({F}, v)$, which satisfies $v=0$ and
\begin{equation}
-  \frac{\rho_{m, end}}{ {{F}_{\text{end}}}^{\frac{3 (w_m+1)}{4}}  } (3 w_m+1)    {F} ^{\frac{3 w_m +1}{4}}  + {{F}^{\frac{1}{4} (3 A+1)} \left(\beta  {F}^{\frac{3}{4} \alpha  (A+1)}+1\right)^{-\frac{\alpha +1}{\alpha }}
   \left(2 \beta  {F}^{\frac{3}{4} \alpha  (A+1)}-3 A-1\right)}=0.
\end{equation}

\subsection{Evolution of normalized energy densities} 
  
Defining 
  \begin{equation}
  \Omega=\frac{{F}}{3H^2},\quad \Omega_m=\frac{\rho_m}{3 H^2}, \label{HubbleNormaLIZED}
  \end{equation}
  such that 
  \begin{equation}
\Omega f({F})+\Omega_{m}=1,\label{eq95}
\end{equation}
and taking ${F}$ as a dynamical variable, we obtain the dynamical system 
\begin{align}
& \frac{d \Omega}{d N}=\Omega  \left(4 F \Omega  f'(F)+2 \Omega  f(F)+(3 w_m+1)  \Omega_{m} -2\right),\\
& \frac{d \Omega_{m}}{d N}=\Omega_m \left(4 F \Omega  f'(F)+2 \Omega  f(F)+(3 w_m+1)
   (\Omega_m-1)\right),\\
 & \frac{d {F}}{d N}=-4 {F},
\end{align}
defined on the invariant surface \eqref{eq95}. 
The above equation can be solved globally for $\Omega_{m}$, and we obtain a 2D dynamical system given by 

\begin{align}
& \frac{d \Omega}{d N}=\Omega  \left(4 {F} \Omega  f'({F})+f({F}) (1-3 w_m) \Omega +3 w_m-1\right),\\
 & \frac{d {F}}{d N}=-4 {F},
\end{align}
Considering the energy condition $\rho_m\geq 0, \quad \rho_{B}\geq 0$, the phase-plane is defined by 
\begin{equation}
\left\{(\Omega,{F})\in\mathbb{R}^2: 0\leq  \Omega f({F})\leq 1, {F}\geq 0\right\}.
\end{equation}
The system admits the equilibrium points (at the finite region of the phase space): 
\begin{enumerate}
\item $(\Omega,{F})=(0,0)$, whose eigenvalues are $\{-4,-1+3 w_m\}$. It corresponds to the FRW matter-dominated solution that it is a sink for $w_m<\frac{1}{3}$  or a saddle for $w_m>\frac{1}{3}$. 
\item $(\Omega,{F})=\left(\frac{1}{f(0)},0\right)$. 
The eigenvalues are $\{-4,1-3 w_m\}$  that it is a saddle for $w_m<\frac{1}{3}$  or a sink for $w_m>\frac{1}{3}$.
\end{enumerate}

 According to the specific form of $f$, we have equilibrium points at the infinite region of the phase space or other structures like periodic solutions.

\begin{figure}[ht!]
    \centering
    \includegraphics[scale=0.55]{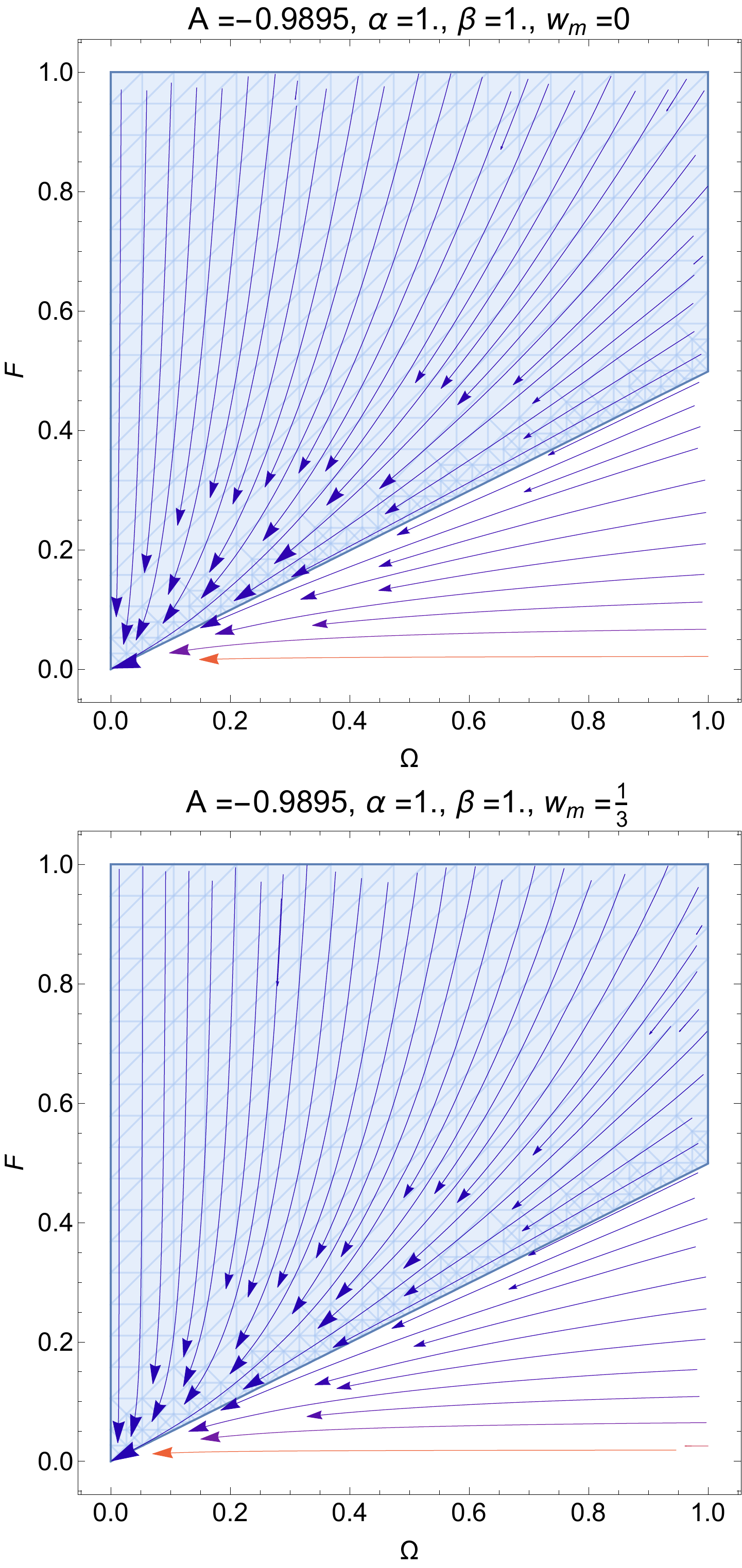}
    \caption{A phase plot of system \eqref{eq110}-\eqref{eq111} for the region of parameters \eqref{Param_3}. The shadowed region correspond to the physical conditions $  0\leq \Omega  {F}^{\frac{1}{4} (3 A-1)} \left(\beta  {F}^{\frac{3}{4} \alpha  (A+1)}+1\right)^{-1/\alpha }\leq 1, {F}\geq 0$.}
    \label{fig:my_labelDS3}
\end{figure}

We apply the procedure to the present model \eqref{eq50}, such that we obtain the dynamical system 
\begin{align}
 &  \frac{d \Omega}{d N}=\Omega  \left(\beta  {F}^{\frac{3}{4} \alpha  (A+1)}+1\right)^{-\frac{\alpha +1}{\alpha }} \Bigg(-3 \beta  (w_m+1) \Omega  {F}^{\frac{3}{4} (\alpha +\alpha  A+A)-\frac{1}{4}}+\beta 
   (3 w_m-1) {F}^{\frac{3}{4} \alpha  (A+1)} \left(\beta  {F}^{\frac{3}{4} \alpha  (A+1)}+1\right)^{\frac{1}{\alpha }}\nonumber \\
   & +(3 w_m-1) \left(\beta  {F}^{\frac{3}{4} \alpha 
   (A+1)}+1\right)^{\frac{1}{\alpha }}+3 \Omega  {F}^{\frac{3 A}{4}-\frac{1}{4}} (A-w_m)\Bigg), \label{eq110}\\
  & \frac{d {F}}{d N}=  -4 {F}. \label{eq111}
\end{align}
defined on the phase space 
\begin{equation}
\left\{(\Omega,{F})\in\mathbb{R}^2:
    0\leq \Omega  {F}^{\frac{1}{4} (3 A-1)} \left(\beta  {F}^{\frac{3}{4} \alpha  (A+1)}+1\right)^{-1/\alpha }\leq 1, {F}\geq 0\right\}.
\end{equation} 
We consider the parameter region 
\begin{align}
& \alpha=1, \quad  A= -0.9895,  \quad  \beta=1,  \quad w_m \in \{0, 1/3\}. \label{Param_3}
\end{align}
In figure \ref{fig:my_labelDS3}, we draw a phase plot of system \eqref{eq110}-\eqref{eq111} for the region of parameters \eqref{Param_3}. The attractor corresponds to the FLRW matter-dominated solution, which is a sink. 

\section{Conclusions}
\label{sec:con}

In the present work, we have investigated the inflation driven by a nonlinear electromagnetic field based on an NLED lagrangian density ${\cal L}_{\text{nled}} = - {F} f \left( {F} \right)$, where $f \left( {F}\right)$ is a general function depending on ${F}$ that encodes nonlinearity. We formulated an $f$-NLED cosmological model with a more general function $f \left( {F}\right)$ and showed that all NLED models could be expressed in this framework; then, we investigate in detail two interesting examples of a function $f \left( {F}\right)$. We presented our phenomenological model based on a new Lagrangian for NLED. Solutions to the field equations with the physical properties of the cosmological parameters were obtained. We have shown that the early Universe had no Big-Bang singularity and tended to accelerate in the past. We also investigate the qualitative implications of NLED by studying the inflationary parameters, like the slow-roll parameters, spectral index $n_s$, and tensor-to-scalar ratio $r$, and compare our results with observational data. Detailed phase-space analysis of our NLED cosmological model was performed with and without matter. We have examined the dynamics of our model by using dynamical systems tools. 

As a first approach, we have considered the motion of a particle of the unit mass in the effective potential. 
This equation is satisfied on the zero-energy level, where $\rho_{B}$ plays the role of effective energy density parameterized through the scale factor $a(t)$. Therefore, the standard cosmological model can be represented in terms of a dynamical system of a Newtonian type under a given potential $W(u)$. Thus,  $\frac{v^2}{2}+W(u)=E$, is the constant of energy. Generically, the equilibrium points of the resulting system are situated on the axis $u$ ($v=0$), and they satisfy $\frac{\partial W(u)}{\partial u}=0$. From the characteristic equation, it follows that just three types of equilibrium points are admitted: 

\begin{enumerate}
\item Saddle if $u_c: \frac{\partial W}{\partial u}|_{u=u_c}=0$ and $\frac{\partial^2 W}{\partial u^2}|_{u=u_c}<0;$ 
\item Focus if $u_c: \frac{\partial W}{\partial u}|_{u=u_c}=0$ and $\frac{\partial^2 W}{\partial u^2}|_{u=u_c}>0;$ 
\item Degenerated critical point if  $u_c: \frac{\partial W}{\partial u}|_{u=u_c}=0$ and $\frac{\partial^2 W}{\partial u^2}|_{u=u_c}=0.$ 
\end{enumerate}
This heuristic analysis determines the dynamics of the slow manifold. Indeed, in the vacuum case, we have obtained the system \eqref{eq89} and \eqref{eq90}. 
Moreover, according to our estimations,  for the theoretical prior $z_{\text{end}}\simeq 10^{-2} z_{GUT}$,   we have $5 \times 10^{23} \mathrm{cm}^{-2}\lesssim {F}_{\text{end}} \lesssim 5 \times 10^{37} \mathrm{cm}^{-2} $.  Then,  we have
$1.17851\times 10^{-20} \mathrm{cm} \lesssim \varepsilon:= \frac{1}{12 \sqrt{{F}_{\text{end}}}}  \lesssim 1.17851\times 10^{-13} \mathrm{cm}$, therefore, we are in the presence of a fast-slow system. 
The system  \eqref{eq89} and \eqref{eq90} gives the dynamics in the fast manifold. That corresponds to the ``horizontal motion'' $v=v_0$ (constant), and \begin{equation}
    {F}(\tau_0) e^{-4 v_0 (\tau -\tau_0)} \label{horizontal}.
\end{equation} This fact is confirmed numerically by  considering the parameter region \eqref{param1}. For vacuum,  the dynamics at the slow manifold are governed by the equilibrium points $({F}, v)$ which satisfies $v=0$ and \begin{equation}\left(\beta  {F}^{\frac{3}{4} \alpha  (A+1)}+1\right)^{-\frac{\alpha +1}{\alpha }}
   \left(2 \beta  {F}^{\frac{3}{4} \alpha  (A+1)}-3 A-1\right)=0. \label{slow1}
   \end{equation}
   Depending on the parameter values, the attractors are $P_0$ or $P_1$. They were analyzed in the coordinates $(u,v)$. Notice that that the axis $({F},v)=(0,v_c)$ is a line of equilibrium points, with eigenvalues $\{0,-4 v_c\}$. 
   
 Analogously, we analyzed the matter case with dust matter ($w_m=0$), and we assumed $\beta=1$. Using $H_{0}=h\,1.08\times 10^{-30}\mathrm{cm}^{-1}$, where  $h =(67.4\pm 0.5)\times 10^{-2}$,  $\Omega_{m0}=0.315\pm 0.007$ and $N_{\text{eff}}=2.99 \pm 0.17$ according to the Planck 2018 results \cite{Planck2018}, and considering the theoretical prior $z_{\text{end}}\simeq 10^{-2} z_{GUT}$, we have 
$4.8\times 10^{17} \mathrm{cm}^{-2} \lesssim \rho_{m, end} \lesssim 5.2 \times 10^{17} \mathrm{cm}^{-2}$.
We select the best-fit value
$\rho_{m, end} \simeq 5.0\times 10^{17}$.  Then,  we have
$1.17851\times 10^{-20} \mathrm{cm} \lesssim \varepsilon:= \frac{1}{12 \sqrt{{F}_{\text{end}}}}  \lesssim 1.17851\times 10^{-13} \mathrm{cm}$, and for $w_m=0$,  we have  $2.65915 \times 10^{-11} \lesssim \frac{\rho_{m, end}}{ {{F}_{\text{end}}}^{\frac{3 (w_m+1)}{4}}  } \lesssim 0.840896$. Therefore, we are in the presence of a fast-slow system.  
As before, \eqref{Neq89} and \eqref{Neq90}
determine the dynamics in the fast manifold. That corresponds to the ``horizontal motion'' $v=v_0$ (constant), and $ {F}(\tau)$ given by \eqref{horizontal}. 
This fact is confirmed numerically by considering the parameter region  \eqref{param2}. The equilibrium governs the dynamics at the slow manifold points $({F}, v)$ which satisfies $v=0$ and \eqref{slow1} is generalized to \begin{equation}
-  \frac{\rho_{m, end}}{ {{F}_{\text{end}}}^{\frac{3 (w_m+1)}{4}}  } (3 w_m+1)    {F} ^{\frac{3 w_m +1}{4}}  + {{F}^{\frac{1}{4} (3 A+1)} \left(\beta  {F}^{\frac{3}{4} \alpha  (A+1)}+1\right)^{-\frac{\alpha +1}{\alpha }}
   \left(2 \beta  {F}^{\frac{3}{4} \alpha  (A+1)}-3 A-1\right)}=0.  \label{slow2}
    \end{equation}
Due to the difficulties in analyzing the fast-slow dynamics and the dependence on the experimental parameters ${{F}_{\text{end}}}, \rho_{m, end}$ etcetera, we have considered alternative Hubble-normalized variables \eqref{HubbleNormaLIZED}. In the general case, the system admitted the equilibrium points (at the finite region of the phase space):  (i) $(\Omega,{F})=(0,0)$, whose eigenvalues are $\{-4,-1+3 w_m\}$. It corresponds to the FLRW matter-dominated solution that it is a sink for $w_m<\frac{1}{3}$  or a saddle for $w_m>\frac{1}{3}$; and (ii)  $(\Omega,{F})=\left(\frac{1}{f(0)},0\right)$. 
The eigenvalues are $\{-4,1-3 w_m\}$  that it is a saddle for $w_m<\frac{1}{3}$  or a sink for $w_m>\frac{1}{3}$. In the particular case of $f({F})$ in \eqref{eq50} we have obtained the system \eqref{eq110} and \eqref{eq111} where the late-time attractor is the FRW matter dominated solution $(\Omega,{F})=(0,0)$. 
The last analysis complemented the analysis of the fast-slow dynamics.

\section*{Conflict of Interest Statement} 

The authors declare to have no conflict of interest.

\section*{Author Contributions}

The authors equally contributed.  

\section*{Data Availability Statement}

This manuscript has no associated data, or the data will not be deposited. [Authors' comment: This is a Theoretical Research Project.]

\section*{Acknowledgments}
HBB gratefully acknowledges the financial support from the University of Sharjah.
The work of HQ was partially supported by UNAM-DGAPA-PAPIIT, Grant No. 114520, and  Conacyt-Mexico, Grant No. A1-S-31269. The research of Genly Leon is funded by Vicerrector\'ia de Investigaci\'on y Desarrollo Tecnol\'ogico (Vridt) at Universidad Cat\'olica del Norte. He also thanks Vridt support through Concurso De Pasantías De Investigación Año 2022, Resolución Vridt N 040/2022 and Resolución Vridt No. 054/2022. He also thanks the support of Núcleo de Investigación Geometría Diferencial y Aplicaciones, Resolución Vridt No. 096/2022. A. Ö. would like to acknowledge the contribution of the COST Action CA21136 - Addressing observational tensions in cosmology with systematics and fundamental physics (CosmoVerse). The authors thank Esteban González for preparing Figure \ref{fig:r-ns}. An anonymous referee is acknowledged for helpful suggestions.

\end{document}